\documentclass[journal]{IEEEtran}

\usepackage{nomencl}
\usepackage{amsmath}
\usepackage{comment}
\usepackage[table]{xcolor}
\usepackage{lipsum} 
\usepackage{gensymb}
\usepackage{bm}
\usepackage{mathtools}
\usepackage{soul}
\usepackage{multirow}
\usepackage{booktabs,colortbl}
\usepackage[utf8]{inputenc}
\usepackage[T1]{fontenc}
\usepackage{balance}
\usepackage{hhline}
\usepackage{graphicx}
\usepackage{array}
\usepackage{subfigure}
\usepackage{color}
\usepackage{relsize}
\usepackage{cite}
\usepackage{comment}
\usepackage{epstopdf}
\usepackage{float}
\usepackage{tablefootnote}
\usepackage{tabularx}
\usepackage{algorithm,algorithmic}

\usepackage[algo2e, ruled, vlined, linesnumbered]{algorithm2e}

\usepackage{smartdiagram}
\usepackage{graphbox}
\usepackage{stfloats}

\usepackage{tikz}
\usetikzlibrary{trees,positioning,shapes,shadows,arrows}
\usetikzlibrary{arrows.meta}
\usetikzlibrary{automata}

\makeatletter
\renewcommand{\ALG@name}{\footnotesize Pseudocode}
\makeatother

\usepackage{array}
\usepackage{eqparbox}

\renewcommand{\baselinestretch}{0.89} 
\graphicspath{{./figures/}}

\newcommand{\hlblue}[1]{{\color{black} #1}}

\begin{document}
	% Define Tikzset
	\tikzset{
		basic/.style  = {draw, text width=4.5cm, drop shadow, rectangle},
		root/.style   = {basic, rounded corners=0.5pt, thin, align=center, fill=white},
		level-2/.style = {basic, rounded corners=0.5pt, thin,align=center, fill=white, text width=4.5cm},
		level-3/.style = {basic, rounded corners=0.5pt, thin, align=left, fill=white, text width= 4.5cm}
	}
	
	%% paper title	
	%\title{A Review of Grid-Forming Inverter-based Wind Turbine Generators}
	
	\title{Grid-Forming Inverter-based Wind Turbine Generators: Comprehensive Review, Comparative Analysis, and Recommendations}
	
	%% Authors
	\author
	{Thai-Thanh Nguyen,~\IEEEmembership{Member,~IEEE,}
		~Tuyen~Vu,~\IEEEmembership{Member,~IEEE,}
		~Sumit~Paudyal,~\IEEEmembership{Member,~IEEE,}
		~Frede~Blaabjerg,~\IEEEmembership{Fellow,~IEEE,}
		~Thanh~Long~Vu,~\IEEEmembership{Member,~IEEE}
		\vspace{-10pt}
		% \thanks{This work is supported by the New York State Energy Research and Development Authority (NYSERDA) - award number 148516.
		\thanks{T. T. Nguyen and T. Vu are with Clarkson University, NY, USA; S. Paudyal is with  Florida International University, FL, USA. F. Blaabjerg is with  Aalborg University,  Denmark. T. L. Vu  is with the Pacific Northwest National Laboratory. Emails: tnguyen@clarkson.edu, tvu@clarkson.edu, spaudyal@fiu.edu, fbl@energy.aau.dk, thanhlong.vu@pnnl.gov. Corresponding Author: T. Vu, Email: tvu@clarkson.edu}
	}
	
	% \markboth{IEEE Trans. Sustainable Energy}%
	% {Shell \MakeLowercase{\textit{et al.}}: Bare Demo of IEEEtran.cls for IEEE Journals}

	\maketitle
	
	\begin{abstract}
		
		High penetration of wind power with conventional grid following controls for inverter-based wind turbine generators (WTGs) weakens the power grid, challenging the power system stability. Grid-forming (GFM) controls are emerging technologies that can address such stability issues. Numerous methodologies of GFM inverters have been developed in the literature; however, their applications for WTGs have not been thoroughly explored. As WTGs need to incorporate multiple control functions to operate reliably in different operational regions, the GFM control should be appropriately developed for the WTGs. This paper presents a review of GFM controls for WTGs, which covers the latest developments in GFM controls and includes multi-loop and single-loop GFM, virtual synchronous machine-based GFM, and virtual inertia control-based GFM. \hlblue{A comparison study for these GFM-based WTGs regarding normal and abnormal operating conditions together with black-start capability is then performed. The control parameters of these GFM types are properly designed and optimized to enable a fair comparison. In addition, the challenges of applying these GFM controls to wind turbines are discussed, which include the impact of DC-link voltage control strategy and the current saturation algorithm on the GFM control performance, black-start capability, and autonomous operation capability. Finally, recommendations and future developments of GFM-based wind turbines to increase the power system reliability are presented.}

	\end{abstract}
	
	
	\IEEEpeerreviewmaketitle

	\section{Introduction}
	
	%\IEEEPARstart{T}{he} need of offshore wind energy system with recent projects. 
	\IEEEPARstart{W}{ind} turbine generators, especially for offshore wind farms, are growing in both size and power rating to reduce the Levelized cost of energy \cite{SHIELDS2021117189}. GE launched the 12-14\,MW Haliade-X offshore turbine in 2021~\cite{GE14MWhaliade}, while Vestas  also introduced 10\,MW offshore turbines in 2018 and 15\,MW offshore wind turbine in 2021~\cite{Vestas10MW,Vestas15MW}, to name a few. In 2020, National Renewable Energy Laboratory (NREL) released the International Energy Agency (IEA) 15~MW offshore reference wind turbine~\cite{gaertner2020iea}. Multi-megawatt wind turbines are typically coupled with the power conversion systems to increase the efficiency of wind turbines and support grids under normal and abnormal operations. Controls of power conversion systems in multi-megawatt wind turbines for large wind farms will play an essential role in improving the power system stability with high wind penetration.
	
	Typical turbine controllers are based on the grid-following (GFL) techniques, in which the phase-locked loop (PLL) is used to track the instantaneous angle of the point of common coupling (PCC) voltage, and the current control loop is used to regulate the AC current reflecting active and reactive power injected into the grid. The main issue in using PLL is that it negatively reduces the power converter stability's margin, especially when the grid condition is weak~\cite{9408354}. \hlblue{Advanced controls of GFL can enhance the stable operation range, but it still requires a minimum level of system strength to maintain stability \cite{ESIG_GFM}.} The instability of multi-megawatt wind turbines will result in more severe impacts on the overall power system stability. Grid-forming (GFM) controls can address the instability issues of wind turbines under weak grid conditions as they do not require the PLL~\cite{9282199}. Instead, a wind turbine with GFM control behaves as a controllable voltage source behind a low-output impedance, which forms the voltage amplitude and frequency of the local grid. Thus, the GFM wind turbines can operate independently from the main power grid and possess the black start capability with a sufficient energy buffer. \hlblue{Successful trial of a 69\,MW GFM wind park and demonstration of Type-3 GFM wind turbine by GE and NREL reveal GFM wind turbines' potentials with additional promising inertia support and black start capability ~\cite{roscoe2020response, roscoe2019practical, GENRELGFM}.}
	
	While the definition of the GFM controls is being considered in industrial and academic communities, various GFM structures for wind turbines have been introduced~\cite{avazov2021application, li2021novel, jain2020grid, yazdi2019analytical}. Although there are existing review studies of GFM inverters \cite{lin2020research, 8932418, 8879610, rokrok2020classification}, they do not cover GFM applications for wind turbines. Since the control systems of wind turbines are complex with multiple operational regions, along with multiple control functions, such as maximum power point control, constant torque/speed control, voltage-ride though control, etc., the uses of GFM controls in wind turbines should be carefully considered to ensure that all functions are preserved.  In this context, this paper aims to contribute  the following.
	\hlblue{
		\begin{itemize}
			\item This paper comprehensively reviews state-of-the-art GFM controls of wind turbines. Recognizing the essential difference between GFM and GFL WTG in regulating the DC-link voltage and the importance of DC-link voltage to WTG's performance, we categorized the GFM WTGs based on the regulation strategies of DC-link voltage.  Challenges and issues in deploying GFM controls for wind turbines are also discussed. 
			\item A comparative study of different types of GFM controls for wind turbines is carried out to evaluate their advantages and disadvantages.
			\item Future directions and recommendations for GFM wind turbines are provided for future grid stability and resilience through enhancements of controls, protection, standards, modeling, storage integration, and testing.
	\end{itemize}}
	
	%	Next section will present preliminary comparison results between GFL control and two types of GFM controls to address the need to understand different GFM control schemes. 
	
	%	\hlblue{This paper is organized as follows. } 
	%\vfill
	
	% \clearpage
	
	The rest of this paper is organized as follows: Section~\ref{sec:typWTG} summarizes the typical control of the GFL WTG.  Section~\ref{sec:GFMWTG} reviews the state-of-the-art GFM controls of wind turbine generators and presents four types of GFM controls used in this paper for the comparative study. The comparative analysis of these GFM controls under normal and abnormal conditions is presented in Section~\ref{sec:comparative analysis}. \hlblue{The sensitivity analysis and optimal parameters design are also presented.} Future direction and recommendations are provided in Section~\ref{sec:Future}. Finally, Section~\ref{sec:Conclusion} summarizes the main findings of this paper. 
	
	%	\clearpage
	% SECTION II
	
	\section{Grid-Following Wind Turbine Generators}
	\label{sec:typWTG}

	\begin{figure}[b]
		\centering
		\vspace{-10pt}
		\includegraphics[width=0.8\linewidth]{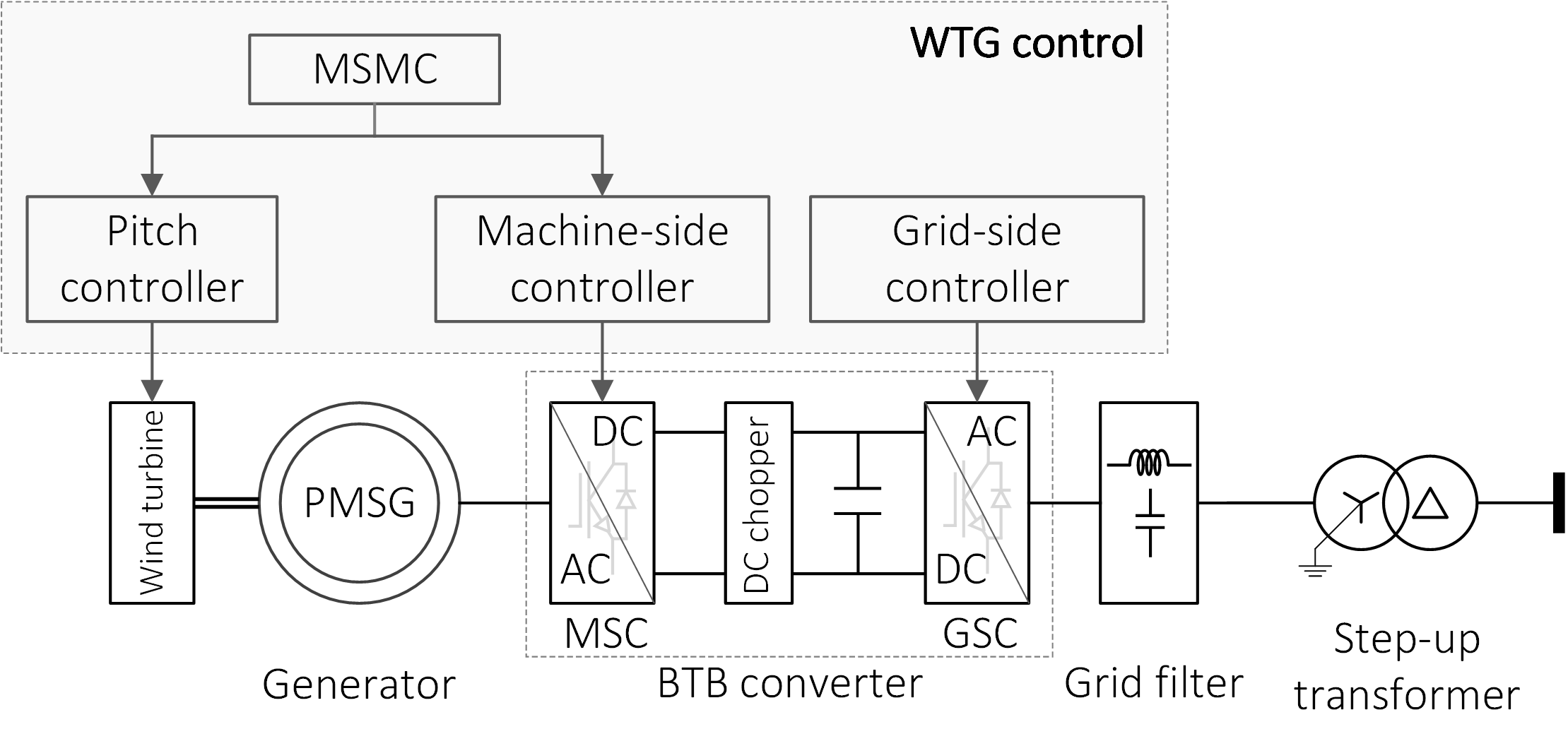}
		\caption{Configuration of Type-4 wind turbine generator and its control system.}
		\label{fig:typ_WTGcontrol}
	\end{figure}
	
	%	\begin{figure*}[bp]
	%		\centering
	%		\includegraphics[width=0.8\linewidth]{GFL.png}
	%		\caption{Typical grid-following control of wind turbine generator.}
	%		\label{fig:GFL}
	%	\end{figure*}
	
	\begin{figure}[!btp]
		\centering
		\includegraphics[width=1.0\linewidth]{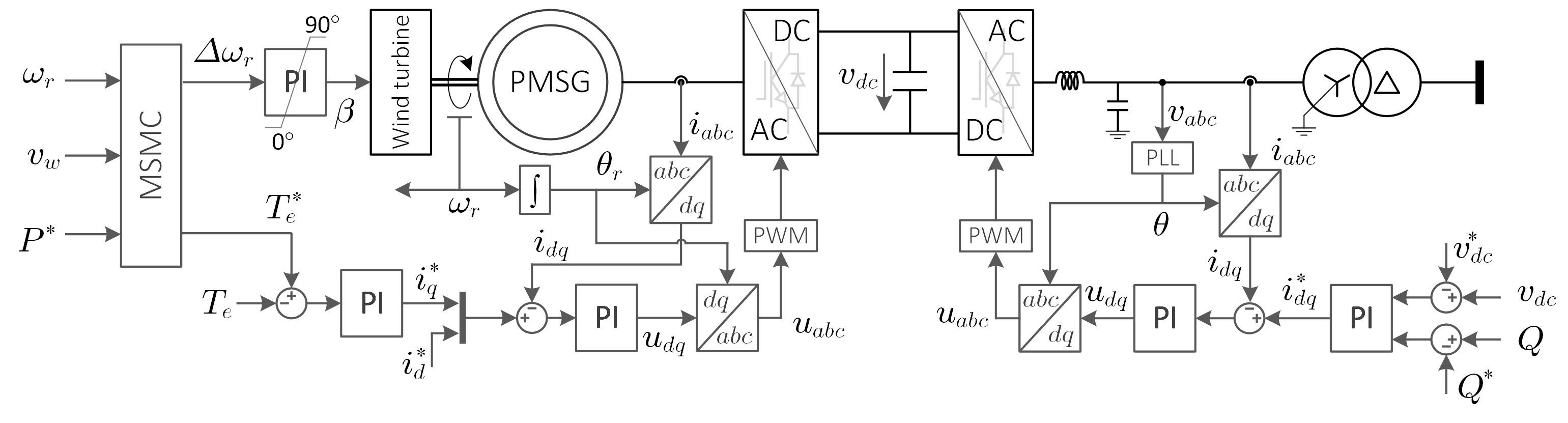}
		\caption{Typical grid-following control of wind turbine generator.}
		\label{fig:GFL}
	\end{figure}
	
	Utility-scale wind turbines are typically based on two types of generators: Type~3 with the doubly fed induction generator (DFIG) and Type~4 with the permanent magnet synchronous generator (PMSG). The former equips with a partial-scale power converter, while the latter equips with a full-scale power converter. Although Type~4 WTG can use other types of generators, such as squirrel cage induction generator (SCIG), wound rotor synchronous generator (WRSG), and high-temperature superconducting synchronous generator (HTS-SG), PMSG-based Type~4 is preferred due to its advantages of high efficiency, high power density and reliability \cite{YARAMASU201737}. PMSG-based WTG offers a variety of advantages compared to the DFIG type. For instance, using a full-scale power converter allows the generator to fully decouple from the grid and provide a wider range of controllability. The gearbox system is used in the PMSG-based WTG to reduce the size and weight of generators. However, due to high maintenance cost and the risk of gearbox failure, the direct-driver PMSG that eliminates the gearbox system from the wind turbines are growing in popularity. As the PMSG-based WTG is becoming the preferred technology \cite{7927779}, this paper mainly focuses on the grid-forming control of this type for review and analysis.
	
	As a majority of controllers in the GFM WTG is the same as the conventional GFL WTG, this section briefly summarizes the GFL control of Type-4 WTG. An overall configuration of Type-4 WTG is shown in Fig.~\ref{fig:typ_WTGcontrol}, which consists of a wind turbine, a permanent magnet synchronous generator (PMSG), a back-to-back (BTB) converter including machine-side converter (MSC), and a grid-side converter (GSC), a DC chopper, grid filters, and a step-up transformer. The main control system includes a pitch controller, a machine-side controller, a grid-side controller, and a machine-side master controller (MSMC). The pitch controller regulates the blade pitch angle in conditions of a high wind speed or curtailing wind power. The grid-side controller is responsible for maintaining the DC-link voltage of the BTB converter, whereas the machine-side controller is in charge of regulating generator torque. The MSMC generates the reference signals for the machine-side and pitch controllers, performing various operation modes.
	
	The detailed schematic diagram of the GFL WTG is shown in Fig.~\ref{fig:GFL} (extension of Fig.~\ref{fig:typ_WTGcontrol}). Both grid-side and machine-side controllers include multi-loop control structure, including an inner current loop and outer DC-link voltage control or torque control loop. The grid-side controller is designed with the grid-following control scheme, in which  PLL is used to track the instantaneous angle of the terminal voltage. The proportional-integral (PI) regulators are used to control voltage and current in $dq$ frame. The rotor speed ($\omega_r$) is used for the machine-side controller to generate the instantaneous phase angle ($\theta_r$). The detailed schematic diagram of grid-side and machine-side converters can be found in \cite{9165711}.
	
	\begin{algorithm}[b]
		\algsetup{linenosize=\footnotesize}
		\algsetup{indent=2em}
		\footnotesize
		\begin{algorithmic}[1]
			\IF [Power curtailment] {$P^* < \bar{P}$} 
			\STATE $\Delta \omega = \omega_r - \bar{\omega}_r$ \\
			\STATE $T_e^* = P^* / \omega_r$
			\ELSE
			\IF[Region 1]{$v_w < v_{\text{c-in}}$} 
			\STATE $\Delta \omega = 0$ \\
			\STATE $T_e^* = 0$
			\ELSIF [Region 1.5] {$v_{\text{c-in}} \leq v_w < v_{\text{inter}} $}
			\STATE $\Delta \omega = \omega - \omega_{\text{min}} $\\
			\STATE $T_e^* = k_{\text{opt}} \omega_r^2$ 
			\ELSIF [Region 2] {$v_{\text{inter}} \leq v_w < \bar{v}_w $}
			\STATE $\Delta \omega = 0 $ \COMMENT{Force $\beta = 0$}\\
			\STATE $T_e^* = k_{\text{opt}} \omega_r^2$
			\ELSIF [Region 3] {$\bar{v}_w \leq v_w < v_{\text{c-out}}$}
			\STATE $\Delta \omega = \omega_r - \bar{\omega}_r $\\
			\STATE $T_e^* = \bar{T}_e$
			\ELSE [Region 4]
			\STATE $\Delta \omega = 0$\\
			\STATE $T_e^* = 0$
			\ENDIF
			\ENDIF 
		\end{algorithmic} 
		\caption{\footnotesize Machine-side master control.}
		\label{alg:MSMC}
	\end{algorithm}
	%	\begin{figure}[t]
	%		\centering
	%		\includegraphics[width=0.9\linewidth]{powercurve.png}
	%		\caption{ Operating regions. Note that regions 1 and 4 are not shown.}
	%		\label{fig:powercurve}
	%	\end{figure}	
	The reference signals for the pitch and machine-side controllers depend on the wind speed condition and the reference power from wind plant-level control. An additional outer control loop generates these reference signals, which is represented by the block diagram in existing studies \cite{9589222, 5262959, 5342521}. However, in this paper MSMC, which has the same functionality as such outer loop, is introduced for the sake of simplicity, as given by \textbf{Pseudocode}~\ref{alg:MSMC}. In power curtailment mode, the pitch controller regulates rotor speed at the rated value ($\bar{\omega}_r$) while the machine-side controller regulates the output wind power. Otherwise, the wind turbine generator operates under four main regions depending on the wind speed condition. The wind turbine is turned off in regions 1 and 4 where the wind speed is lower than the cut-in speed ($v_{\text{c-in}}$) or higher than the cut-out speed ($v_{\text{c-out}}$), respectively. To avoid 3-period interference effects, the pitch controller regulates rotor speed at minimal value ($\omega_{\text{min}}$) in Region 1.5 \cite{gaertner2020iea} where the wind speed is higher than the cut-in speed but lower than the wind speed caused interference ($v_{\text{inter}}$). In Region 2, where the wind speed is higher than the $v_{\text{inter}}$ but lower than the rated wind speed ($\bar{v}_w$), the pitch controller is disabled, and the generator torque is regulated optimally to maximize the wind power, in which the optimal gain ($k_{\text{opt}}$) is given by (\ref{eq:kopt}). 
	\begin{align}
		\label{eq:kopt}
		k_{\text{opt}}=0.5 \rho \pi R^2 C_{p}^{\text{max}} (R / \lambda_{\text{opt}})^3.
	\end{align}
	where $\rho$ is the air density; $R$ is the turbine rotor radius; $C_{p}^{\text{max}}$ is the maximal rotor power coefficient; and $\lambda_{\text{opt}}$ is the optimal tip-speed ratio.
	When the wind speed is higher than the rated wind speed but lower than the cut-out speed, the wind turbine operates on Region 3. In this region, the pitch controller is activated to regulate the rotor speed and generator torque at the rated values ($\bar{\omega}_r$ and $\bar{T}_e$). 
	%	\begin{figure}[t]
	%		\centering
	%		\includegraphics[width=0.9\linewidth]{powercurve.png}
	%		\caption{ Operating regions. Note that regions 1 and 4 are not shown.}
	%		\label{fig:powercurve}
	%	\end{figure}
	
	\section{Grid-Forming Control of WTG}
	\label{sec:GFMWTG}
	
	\begin{figure*}[t]
		\centering
		\begin{tikzpicture}[
			level 1/.style={sibling distance=17em, level distance=3.5em},
			level 2/.style={sibling distance=5cm},
			%level 3/.append style
			%{edge from parent fork down},
			edge from parent/.style={->,solid,black,thick,sloped,draw}, 
			edge from parent path={(\tikzparentnode.south) -- (\tikzchildnode.north)},
			>=latex, node distance=1.2cm, edge from parent fork down,
			font=\footnotesize]
			
			% root of the the initial tree, level 1
			\node[root] {\textbf{GFM controls of Type-4 Wind Turbine Generators}}
			% The first level, as children of the initial tree
			child {node[level-2] (c1) {\textbf{G-GFM}: GSC controls DC voltage}}
			child {node[level-2] (c2) {\textbf{M-GFM}: MSC controls DC voltage}}
			child {node[level-2] (c3) {\textbf{E-GFM}: ESS controls DC voltage}};
			
			% The second level, relatively positioned nodes
			\begin{scope}[every node/.style={level-3}]
				\node [below of = c1, xshift=30pt] (c11) 
				{\textbf{G-MGFM} \\
					Virtual inertia control (VIC) \cite{li2016advanced}
				};
				\node [below of = c11, yshift=-5pt] (c12) 
				{\textbf{G-SGFM} \\
					VIC \cite{he2018coordinated}, inertia synchronous control (ISynC) \cite{sang2019control},	power synchronous control (PSC) \cite{li2021novel}
				};
				
				\node [below of = c2, xshift=30pt] (c21) 
				{\textbf{M-MGFM} \\
					VIC \cite{wang2015control}
				};
				\node [below of = c21, yshift=-5pt] (c22) 
				{\textbf{M-SGFM} \\
					Virtual synchronous machine (VSM) \cite{avazov2021application,yazdi2019analytical, xi2017inertial, duckwitz2014synchronous}, synchronverter \cite{zhong2015grid}
				};

				\node [below of = c3, xshift=30pt] (c31) 
				{\textbf{E-MGFM} \\
					Virtual synchronous generator (VSG) \cite{ma2017virtual,ma2016voltage}
				};
				\node [below of = c31, yshift=-5pt] (c32) 
				{\textbf{E-SGFM} \\
					VSG \cite{liu2018capacitor}
				};
				
			\end{scope}
			
			% lines from each level 1 node to every one of its "children"
			\foreach \value in {1,2}
			\draw[->] (c1.188) |- (c1\value.west);
			
			%					\foreach \value in {1,...,3}
			\foreach \value in {1,2}
			\draw[->] (c2.188) |- (c2\value.west);
			
			\foreach \value in {1,2}
			\draw[->] (c3.188) |- (c3\value.west);
		\end{tikzpicture}
		\caption{Existing grid-forming controls of Type-4 wind turbine generators.}
		\label{fig:GFMreview}
	\end{figure*}
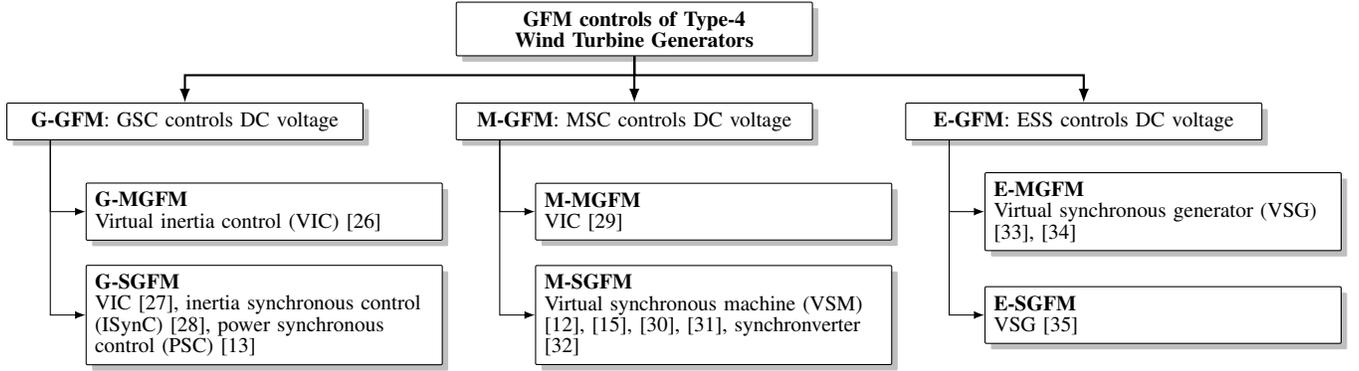
	
	Unlike the GFL WTG, the DC-link voltage of the GFM WTG can be regulated by either grid-side, machine-side, or external converters. Existing studies categorized grid-forming inverters based on the GFM control methodologies, which is appropriate as the DC-link voltage is assumed to be constant. However, for wind turbine applications, the strategy of DC-link voltage regulation must be considered because it affects the implementation of WTG's control system. This paper categorizes the GFM WTGs based on the regulation strategies of DC-link voltage, as shown in Fig.~\ref{fig:GFMreview}. The GFM WTGs are classified into three categories: 
	\begin{itemize}
		\item G-GFM: GSC controls DC voltage.
		\item M-GFM: MSC controls DC voltage.
		\item E-GFM: External energy storage controls DC voltage. 
	\end{itemize}
	In each category, two types of GFM controls are classified according to the inner control loop of the grid-forming controller: the multi-loop control (MGFM) and single-loop control (SGFM). The MGFM types include the inner current and AC voltage control loops, while the SGFM types consist of only the AC voltage control loop. 
	
	The first category of GFM-WTG is the G-GFM, in which the grid-side converter controls the DC-link voltage. An outer DC-link regulator is designed in addition to the inner control loop of the grid-side converter. The DC-link voltage is controlled by adjusting the instantaneous phase angle of the terminal voltage. The DC-link voltage regulators mimic the inertia support of the synchronous generator by allowing the variation of DC-link voltage in an acceptable range, which has been presented in different names, such as virtual inertia control (VIC) \cite{li2016advanced, he2018coordinated}, inertia synchronous control (ISynC) \cite{sang2019control}, and power synchronous control \cite{li2021novel}. The pitch and machine-side controllers are kept the same as the GFL type. 
	
	The second category of GFM WTG is the M-GFM, in which the machine-side converter regulates the DC-link voltage, whereas the grid-side converter is designed for managing output power. The grid-forming controllers implemented in the grid-side converter also mimic the inertia characteristic of synchronous generators, although different controller names have been presented, such as VIC \cite{wang2015control}, virtual synchronous machine (VSM) \cite{avazov2021application,yazdi2019analytical, xi2017inertial, duckwitz2014synchronous}, and synchronverter \cite{zhong2015grid}. 
	
	The last category of GFM WTG is the E-GFM, in which the DC-link voltage is controlled by an external energy storage system (ESS). This approach provides an additional degree of freedom for grid-side controllers while retaining all control functions of pitch and machine-side controllers. VSG-based GFM controls are mainly used in this type \cite{ma2017virtual,ma2016voltage, liu2018capacitor}. As the DC-link voltage is managed constantly by the external ESS, most exiting GFM methodologies available in the literature can be used for the grid-side converter without any modification, such as droop control, power synchronization control, and virtual synchronous generator \cite{9541070, 9408354, 9513281}. It is anticipated that using additional ESS devices introduces technical benefits because ESS can play the role of energy buffer to mitigate the fluctuations in wind power or support grid during the disturbance. However, this GFM type increases the complexity of the WTG control system and total investment cost. 
	
	Overall, it can be found that all existing GFM control methodologies for WTGs try to mimic the inertia characteristic of synchronous generators. This paper investigates only G-GFM and M-GFM categories as they are potential solutions for developing GFM WTG from the existing GFL type. Among two categories, four types of GFM WTGs will be presented in the following sections, which are:
	\begin{itemize}
		\item G-GFM with multi-loop control
		(G\nobreakdash-MGFM)
		\item G-GFM with single-loop control (G\nobreakdash-SGFM)
		\item M-GFM with multi-loop control (M\nobreakdash-MGFM)
		\item M-GFM with single-loop control (M\nobreakdash-SGFM)
	\end{itemize}	
	
	To ensure a fair comparison, the G-MGFM and G-SGFM types use the same outer VIC scheme, while M-MGFM and M-SGFM types use the same outer VSM scheme. The blue color depicts the difference between the GFM and GFL controls of WTGs in the schematic diagram in the following sections. The DC chopper and reactive power control loop of GSC are omitted in the explanation as they are the same for all GFM and GFL control methods. 
	
	\subsection{\textbf{G-GFM:} Grid-side converter controls DC link voltage}
	
	Like the GFL control of WTG, the GSC in the G-GFM category is responsible for regulating DC-link voltage. However, the grid-forming control scheme is used for the grid-side controller. The VIC scheme controls the DC-link voltage, which generates the instantaneous phase angle reference for the inner control loop. According to the structure of inner control loop, this control approach is classified into multi-loop control type (G-MGFM) \cite{li2016advanced} and single-loop control type (G-SGFM) \cite{he2018coordinated, sang2019control, li2021novel}. 	
	
	\begin{figure}[!bt]
		\centering
		\includegraphics[width=0.9\linewidth]{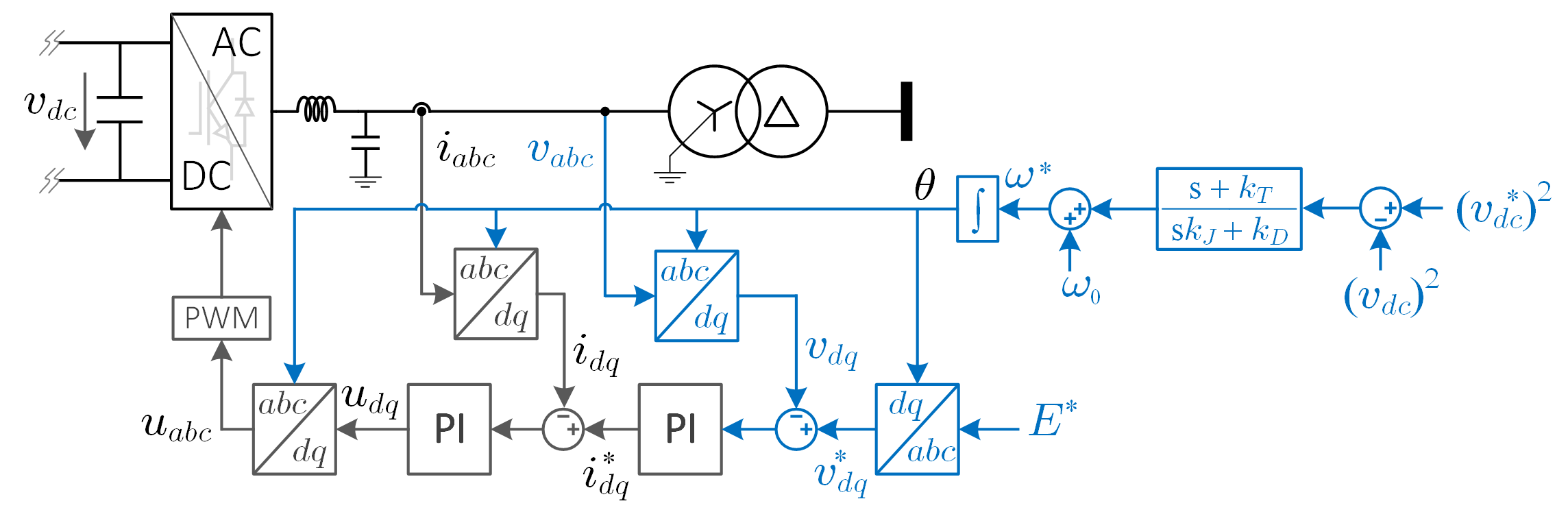}
		\caption{Schematic diagram of G-MGFM wind turbine generator.}
		\label{fig:G-MGFM}
	\end{figure}
	\begin{figure}[!bt]
		\centering
		\includegraphics[width=0.7\linewidth]{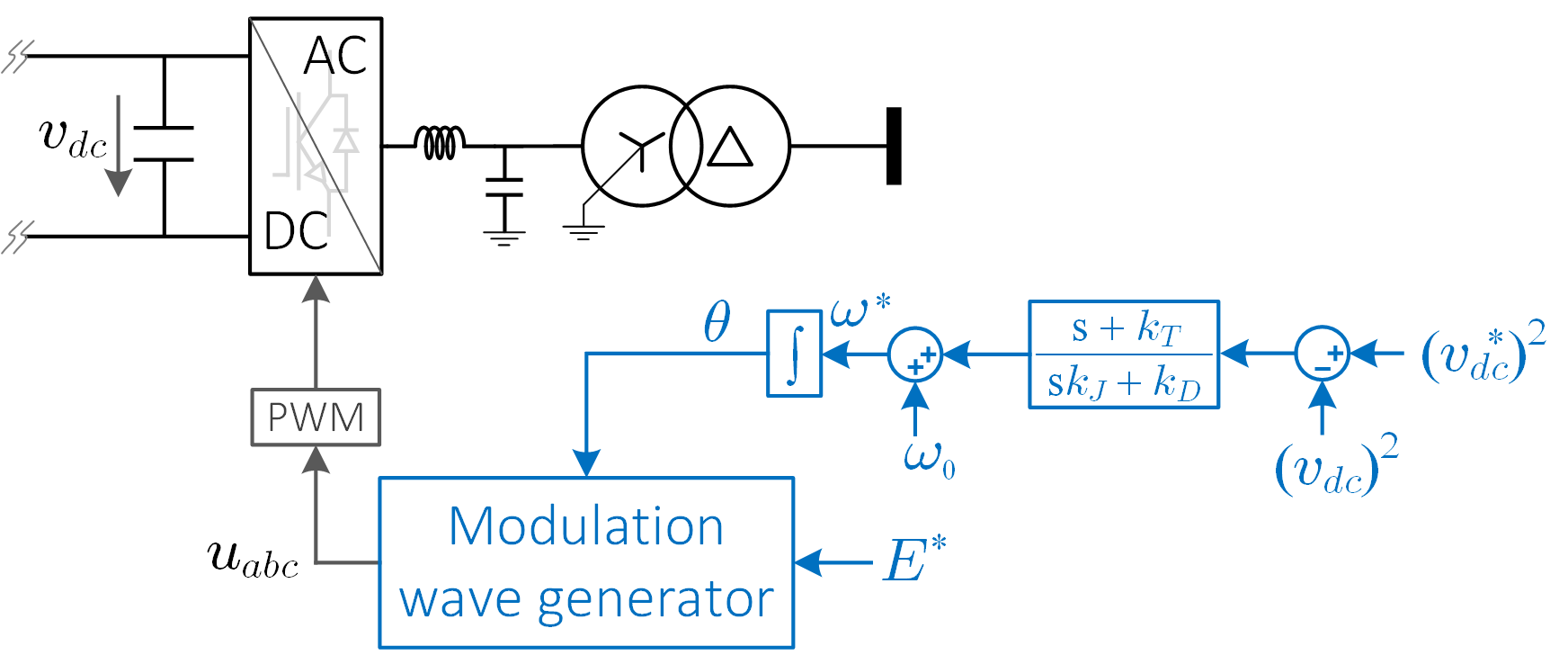}
		\caption{Schematic diagram of G-SGFM wind turbine generator.}
		\label{fig:G-SGFM}
	\end{figure}
	
	The schematic diagram of the G-MGFM control of WTG is shown in Fig.~\ref{fig:G-MGFM}, where the difference between G-MGFM and GFL controls is shown in the blue color. The machine-side controller of the G-MGFM WTG is the same as the GFL WTG, which is omitted in this figure. The grid-side controller utilizes a cascaded control structure that includes an inner current control loop and outer AC voltage control loop, which plays a role in forming the terminal voltage \cite{8863360}. The instantaneous phase angle is directly generated from the reference angular frequency rather than the use of PLL. The angular frequency reference is adjusted to regulate the DC-link voltage of the BTB converter by the virtual synchronous control loop in \cite{li2021novel}, as given by (\ref{eq:theta G-MGFM}). The virtual inertia control utilizes the DC-link as an energy buffer to mimic the behavior of synchronous machine. 
	%	\begin{align}
	%		\label{eq:theta G-MGFM}
	%		\omega^* = \omega_0 -  \Big(k + {1 \over {J\text{s} + D}}\Big) (v_{dc}^{*2} - v_{dc}^2),
	%	\end{align}
	\begin{align}
		\label{eq:theta G-MGFM}
		\omega^* = \omega_0 +  \Big({\text{s} + k_T \over {\text{s}k_J + k_D}}\Big) (v_{dc}^{*2} - v_{dc}^2),
	\end{align}
	where $\omega_0 = 2 \pi f$; $k_T$ is the tracking coefficient; $k_J$ is the inertia coefficient; and $k_D$ is the damping coefficient. 
	
	The schematic control diagram of the G-SGFM is shown in Fig.~\ref{fig:G-SGFM}, in which the modulation signal ($u_{abc}$) is generated directly from the instantaneous phase angle ($\theta$) and reference of voltage amplitude ($E^*$). The modulation wave generator is given by (\ref{eq:Mod-Gen}).
	\begin{align}
		\label{eq:Mod-Gen}
		u_{abc} = E^*
		\begin{bmatrix}
			\sin(\theta) & \sin(\theta - {{2\pi} \over 3}) & \sin(\theta + {{2\pi} \over 3})
		\end{bmatrix}^\top.
	\end{align}
	
	As the role of the grid-side controller is the same as GFL WTG, the pitch and machine-side controllers, including MSMC, are the same as GFL WTG. Thus, all control functions such as region controls and power curtailment are retained.
	
	\subsection{\textbf{M-GFM:} Machine-side converter controls DC link voltage}
	\begin{figure}[tp]
		\centering
		\includegraphics[width=1.0\linewidth]{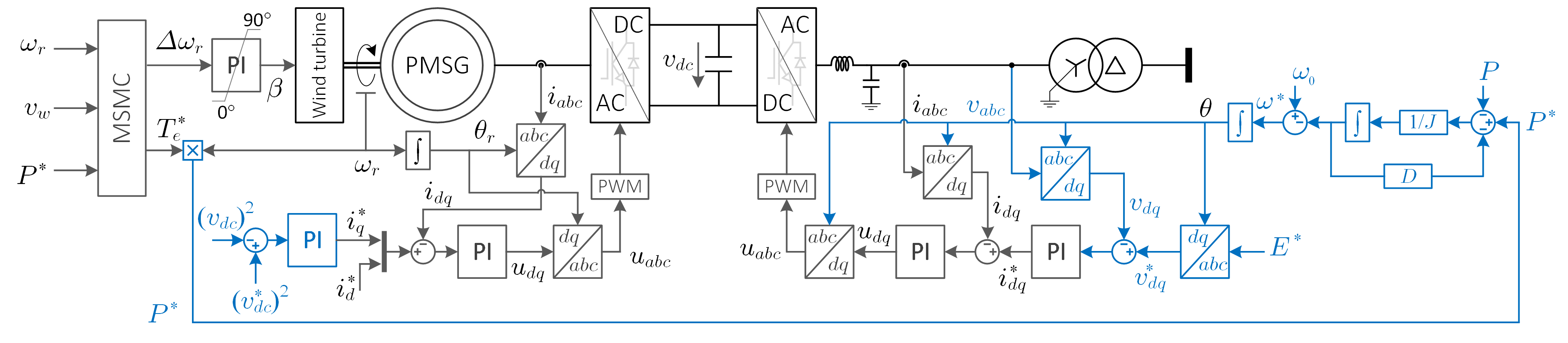}
		\caption{Schematic diagram of M-MGFM wind turbine generator.}
		\label{fig:M-MGFM}
	\end{figure}
	%	\begin{figure*}[tp]
	%		\centering
	%		\includegraphics[width=0.68\linewidth]{M-SGFM.png}
	%		\caption{Schematic diagram of M-SGFM wind turbine generator.}
	%		\label{fig:M-SGFM}
	%	\end{figure*}
	
	%	\begin{figure}[!btp]
	%		\centering
	%		\includegraphics[width=1.0\linewidth]{M-MGFM.png}
	%		\caption{Schematic diagram of M-MGFM wind turbine generator.}
	%		\label{fig:M-MGFM}
	%	\end{figure}
	\begin{figure}[!btp]
		\centering
		\includegraphics[width=1.0\linewidth]{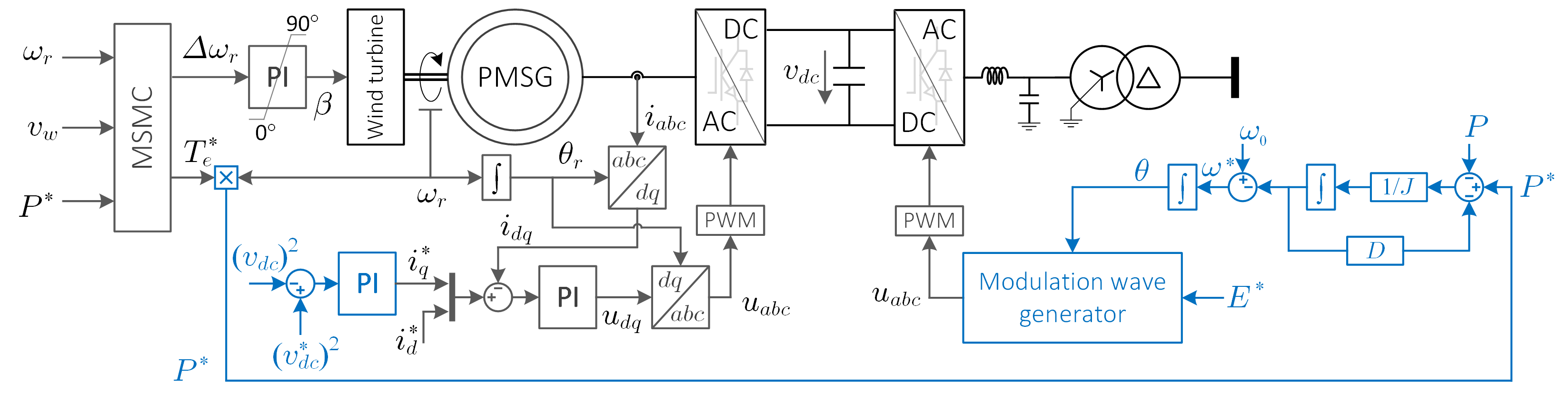}
		\caption{Schematic diagram of M-SGFM wind turbine generator.}
		\label{fig:M-SGFM}
	\end{figure}
	
	The schematic diagrams of the M-MGFM and M-SGFM types are shown in Figs.~\ref{fig:M-MGFM} and \ref{fig:M-SGFM}, respectively. Unlike the GFL WTG, the DC-link voltage of the BTB converter is controlled by the machine-side converter, while the grid-side controller is in charge of regulating the output power of the wind turbine generator. As the control strategies of grid-side and machine-side converters are switched, control in region 2 is modified to retain the control functions of GFL WTG. Instead of optimal torque control to maximize wind power in region 2, the M-GFM manages optimal power ($P^*$) that is calculated by multiplying optimal torque by rotor speed, as given by (\ref{eq:Popt}). The MSMC is the same as GFL WTG.
	\begin{align}
		\label{eq:Popt}
		P^* =  T_e^* \omega_r.
	\end{align}
	
	The virtual synchronous machine (VSM) scheme in \cite{avazov2021application} is used for the outer control loop in this M-GFM category. The reference of angular frequency is adjusted to manage the output power of WTG, as given by (\ref{eq:theta M-MGFM}).
	\begin{align}
		\label{eq:theta M-MGFM}
		\omega^* =  \omega_0 - {1 \over {J\text{s} + D}} (P^* - P),
	\end{align}
	where $P$ is the measured output power of wind turbine generator.

	The cascaded control structure with inner current and voltage loops is used in the M-MGFM type \cite{wang2015control} whereas only voltage control loop is used in the M-SGFM type  \cite{avazov2021application,yazdi2019analytical, xi2017inertial, duckwitz2014synchronous, zhong2015grid}. The control diagram of the M-SGFM wind turbine generator is shown in Fig.~\ref{fig:M-SGFM}, which is the same as M-MGFM control, except the inner controller of the grid-side converter. To generate modulation signals ($u_{abc}$), the modulation wave generator from~(\ref{eq:Mod-Gen}) is used in the M-SGFM type. 	
	
	\section{Comparative Analysis}
	\label{sec:comparative analysis}
	
	%	\begin{table}[!bt]
	%		\renewcommand{\arraystretch}{1.3}
	%		\caption{Parameters of 15MW wind turbine model.}
	%		\label{table:WTGparameters}
	%		\centering
	%		\begin{tabular}{c | c}
	%			\hline\hline
	%			Components & Parameters \\ 
	%			\arrayrulecolor{gray}\hline
	%			\multirow{3}{*}{Step-up transformer} 
	%			& Rated power: 18 MVA \\
	%			& Rated voltage (RMS line): 4 $\mid$ 66 kV \\
	%			& Leakage reactance: 0.1 pu \\
	%			
	%			\hline
	%			
	%			\multirow{5}{*}{BTB converter}
	%			& Rated power: 15 MW\\ 	 
	%			& DC link voltage: 10 kV\\ 	
	%			& DC link capacitance: 4000 $\mu$F\\	  
	%			& Output $LC$ filter: 0.275 mH; 1024 $\mu$F\\
	%			& Switching frequency: 3 kHz\\
	%			
	%			\hline
	%			
	%			\multirow{7}{*}{PMSG} 
	%			& Rated power, $\bar{P}$: 15 MVA\\ 	
	%			& Number of pole pairs, $p$: 162\\	  
	%			& Stator resistance, $R_s$: 0.0368 $\ohm$\\
	%			& Stator $d$-axis inductance, $L_d$: 0.0087 H\\
	%			& Stator $q$-axis inductance, $L_q$: 0.0058 H\\
	%			& Rated speed, $\bar{\omega}_r$: 0.8 rad/s\\
	%			& Total reflected inertia, $J$: $3.16 \times 10^8$ kgm$^2$\\
	%			
	%			\hline
	%			
	%			\multirow{2}{*}{Wind turbine} 
	%			& Rated wind speed, $\bar{v}$: 12~m/s\\ 	
	%			& Turbine rotor diameter, $D$: 236 m\\
	%			
	%			\arrayrulecolor{black}\hline\hline
	%		\end{tabular}
	%	\end{table}

	This section discusses comparative results for five control strategies of WTG: GFL control and four types of GFM controls. A 15~MW direct-drive Type-4 WTG is used to evaluate the control performance of these control strategies. The parameter of 15~MW WTG is given in \cite{9589222}. Firstly, the operation of WTG under the condition of dynamic-wind speed is conducted to evaluate the performance of the region control function. Secondly, under high wind speed conditions, the power set-point of WTG is changed to test the function of power curtailment. Finally, the three-phase-to-ground fault is used to evaluate the responses of these control strategies under abnormal conditions. For all the above scenarios, the grid's strength is also taken into consideration. 
	
	%It should be noted that the parameters of VIC and VSM of all GFM types are the same to ensure a fair comparison in this paper. However, these parameters can be properly designed to obtain the desired performance. 
	
	\hlblue{
		\subsection{Sensitivity Analysis and Optimizing Control Parameters}
		
		The integral and proportional gains of voltage and current control loops were designed using the pole-zero placement approach. A detailed design instruction was presented in \cite{4118327}. The design of the VSM's parameters is the primary focus of this section as it significantly impacts the performance of different GFM-WTG types. The sensitivity analysis is then performed to evaluate the influence of VSM's parameters on the GFM-WTG performance. The gradient descent is used to optimize such parameters in order to achieve the best performance. 
		
		A certain range of each parameter is defined for sensitivity analysis. A given set of parameters is manually tuned and selected using the following steps: 
		\begin{itemize}
			\item Use the trial and error method to find the values of VSM's parameters that maintain the system stability. 
			\item From the stable operation point, gradually decrease and increase the value of each parameter to the point that causes undesired performance in order to find the lower and upper limits of such parameter. This was performed under the rated condition of wind turbines.
			\item Finally, create a set of parameters from the boundaries (upper and lower limits found in the previous step for sensitivity analysis).
		\end{itemize}
		
		The coupling between output active power and DC link voltage is important in designing the VSM's parameters. Three objectives defined in (\ref{eq:SA_req}) are used to evaluate the influence of control parameters on the performance of the GFM WTGs: (\ref{eq:SA_req_1}) represents the deviation of the DC link voltage from the steady-state value; (\ref{eq:SA_req_2}) and (\ref{eq:SA_req_3}) are the peak value and rise time of output active power, respectively. Using Monte Carlo simulations, the sensitivity analysis is performed by evaluating these objectives at each combination of VSM's parameters.
		\begin{subequations}
			\label{eq:SA_req}
			\begin{align}
				\label{eq:SA_req_1}
				\Delta \bar{v}_{dc} & = \int_{t = 0}^{\infty}{|v_{dc}(t)-v_{dc}^{*}| dt}, \\
				\label{eq:SA_req_2}
				P_{\text{max}} & = \max{\{|P(t) - P_{int}|\}}_{t = 0}^{\infty}, \\
				\label{eq:SA_req_3}
				t_{r}(P) & = t_{2} - t_1, 
			\end{align}
		\end{subequations}
		where $v_{dc}(t)$ is the measured DC link voltage at time $t$; $P(t)$ is the measured output power of WTG at time $t$; $P_{int}$ is the initial output power; $t_1$ is the when the output power reaches to 10\% of the steady-state value; $t_2$ is when the output power reaches to 90\% of the steady-state value. 
		
		The optimal performance is achieved by finding the VSM's parameters that minimize the above three objectives. A constrained optimization problem is developed to find the optimal VSM parameters, in which the VSM's parameters are constrained by the lower and upper limits defined in the sensitivity analysis section. In such an optimization problem, the wind turbine model is iteratively simulated. The signals of DC voltage and output active power are recorded to compute the objective values of (\ref{eq:SA_req}). A gradient descent method using a sequential quadratic programming algorithm is used to tune the VSM's parameters to minimize the objectives. Table \ref{table:Opt_VSM} shows the range of each parameter and its optimal value. 
		
		\begin{table}[!btp]
			\renewcommand{\arraystretch}{1.3}
			\caption{Optimal VSM's Parameters.}
			\label{table:Opt_VSM}
			\centering
			\begin{tabular}{c | c | c | c}
				\hline\hline
				GFM Type & Parameter & Range (pu) & Optimal Value (pu) \\ 
				\arrayrulecolor{gray}\hline
				\multirow{3}{*}{G-MGFM} 
				& $k_T$		& [1.0 60.0]		& 41.89 \\	
				& $k_J$		& [0.001 0.1]		& 0.029 \\
				& $k_D$		& [1.0 50.0]		& 3.70 \\			
				\hline
				
				\multirow{3}{*}{G-SGFM} 
				& $k_T$		& [0.1 60.0]		& 55.04 \\	
				& $k_J$		& [0.001 0.1]		& 0.053 \\
				& $k_D$		& [1.0 50.0]		& 7.07 \\			
				\hline
				
				\multirow{2}{*}{M-MGFM} 	
				& $J$		& [0.1 1.0]		& 0.608 \\
				& $D$		& [2.0 10.0]		& 5.080 \\			
				\hline
				
				\multirow{2}{*}{M-SGFM} 	
				& $J$		& [0.1 1.0]		& 0.3 \\
				& $D$		& [2.0 10.0]		& 2.9 \\			
				\hline
				
				\arrayrulecolor{black}\hline\hline
			\end{tabular}
		\end{table}
		
		Fig.~\ref{fig:SA_GFM} depicts the GFM wind turbine performance with the optimal and random parameters. Figs.~\ref{fig:SA_GFM}(a) and (c) show the performance of the G-GFM types, and Figs.~\ref{fig:SA_GFM}(b) and (d) depict the performance of the M-GFM types. It should be noted that the rise time of active power indicates the inertia support for the grid. The variation of VSM's parameters in the G-GFM types has no impact on the rise time of active power response. It is reasonable as the VSM in the G-GFM types is used for DC-link voltage regulation. Thus, the machine-side controller should be improved in order to provide inertia support for the grid. By comparison, the VSM's parameters of M-GFM types have a significant impact on the active power response. Therefore, the amount of inertia support can be chosen by tuning such parameters. The coupling effect between the DC-link voltage and active output power is clearly shown in Figs.~\ref{fig:SA_GFM}(b) and (d). The fast response in active power gives rise to a significant oscillation of DC-link voltage. The proposed optimization problem can find the optimal VSM parameters to balance the trade-off between the DC link voltage oscillation and active power response. 
		
		\begin{figure}[!bt]
			\centering
			\subfigure[G-MGFM]
			{
				\includegraphics[width=0.46\linewidth]{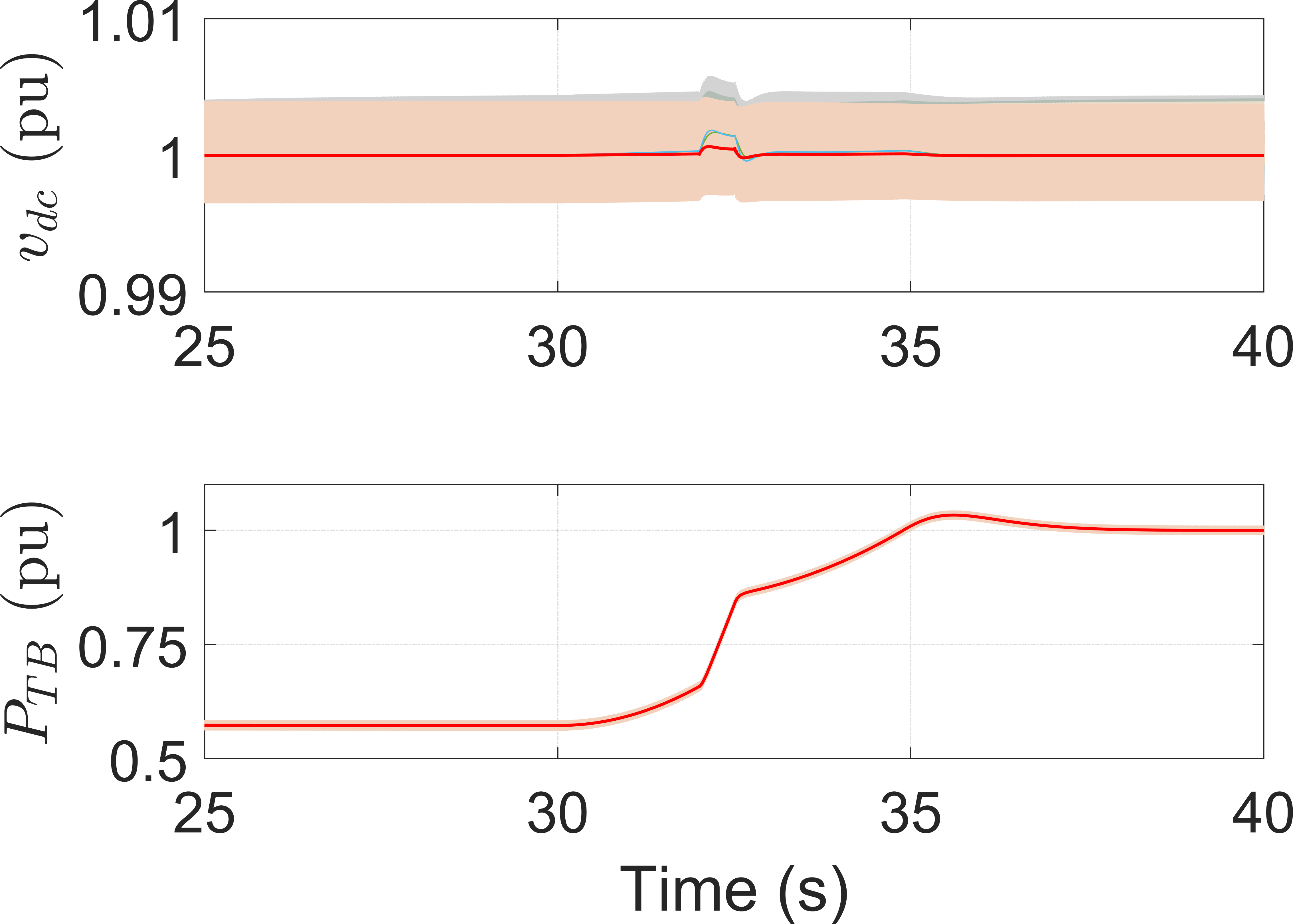}
			} 
			\subfigure[M-MGFM]
			{
				\includegraphics[width=0.46\linewidth]{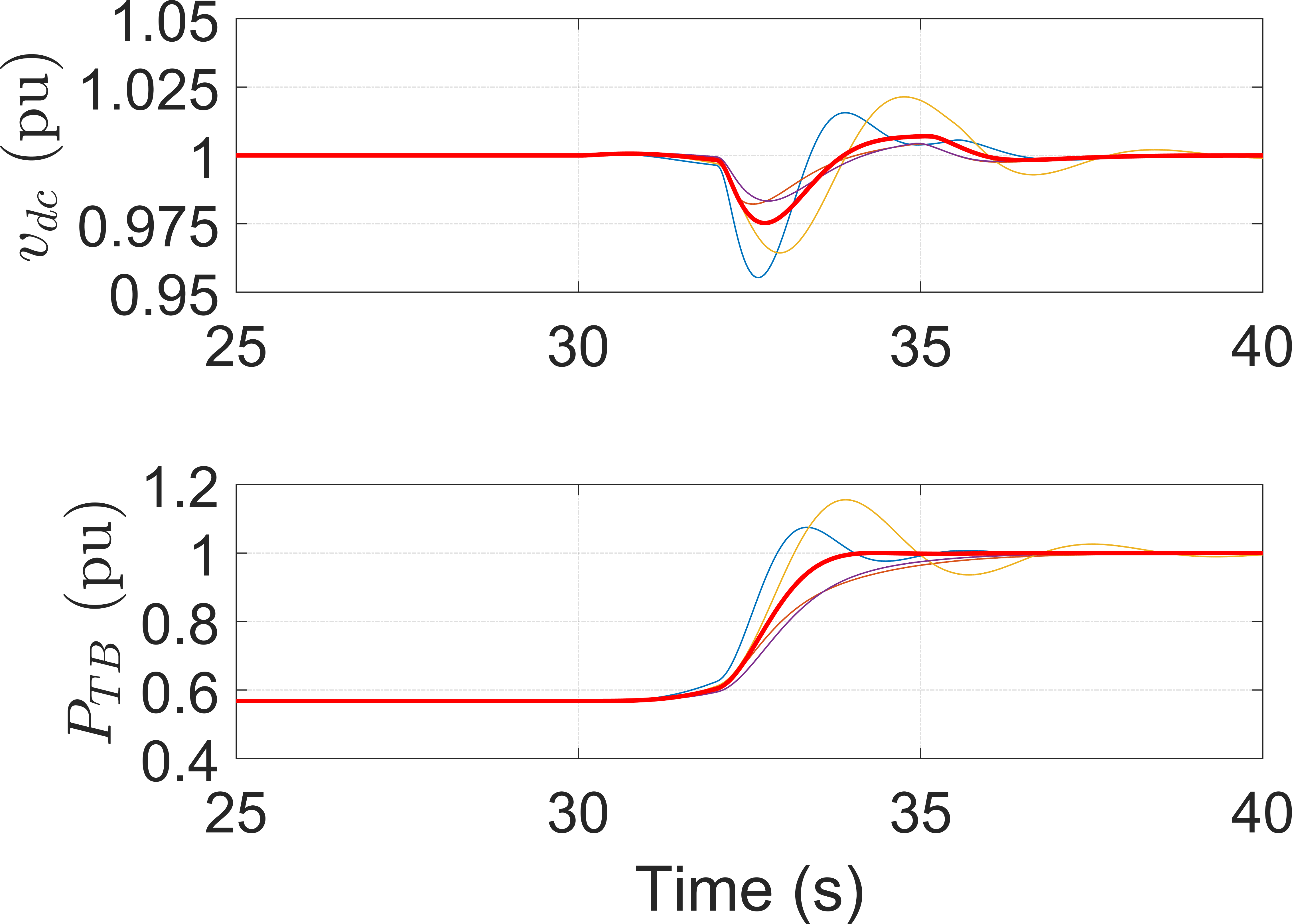}
			} 
			
			\subfigure[G-SGFM]
			{
				\includegraphics[width=0.46\linewidth]{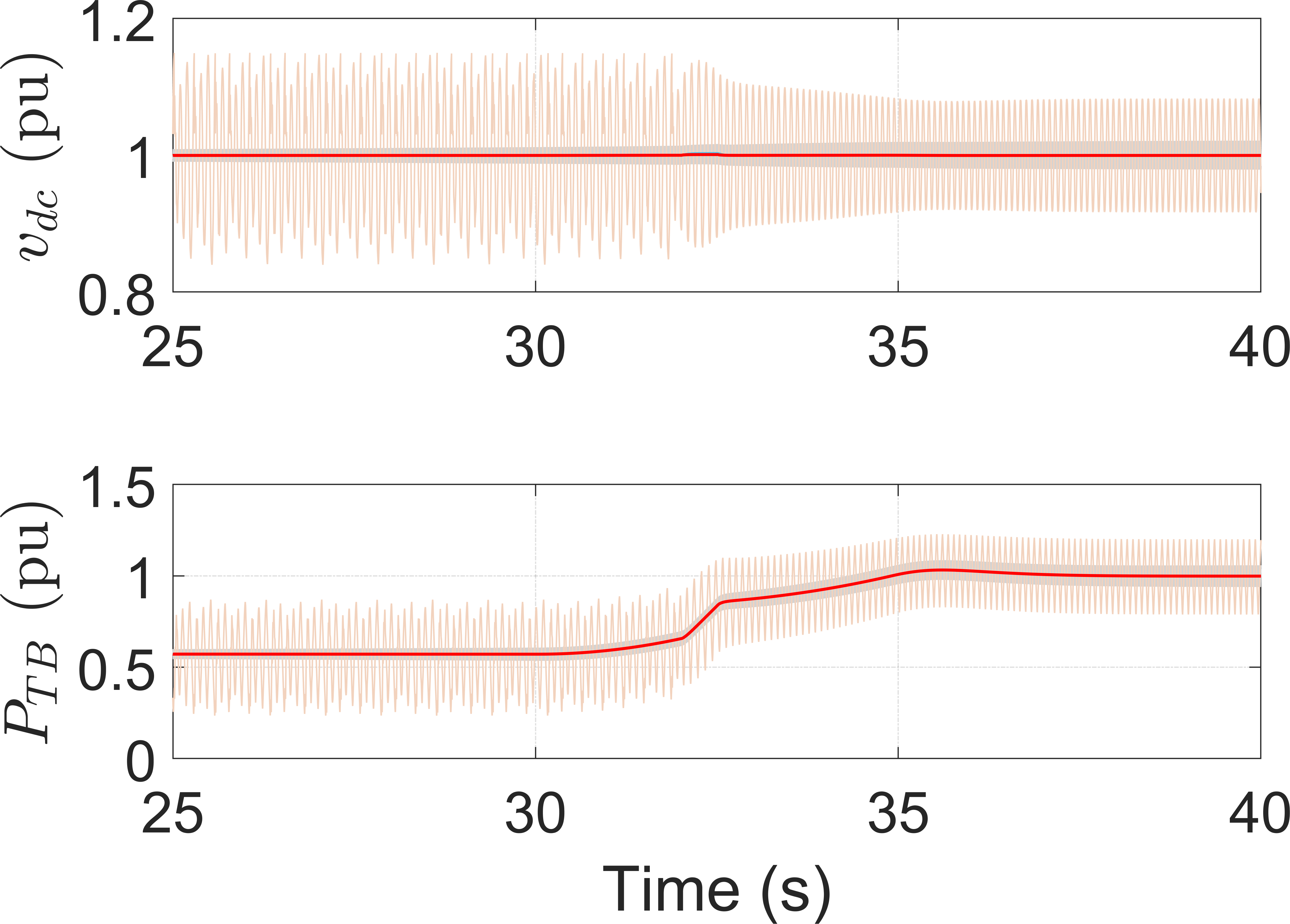}
			} %\quad
			\subfigure[M-SGFM]
			{
				\includegraphics[width=0.46\linewidth]{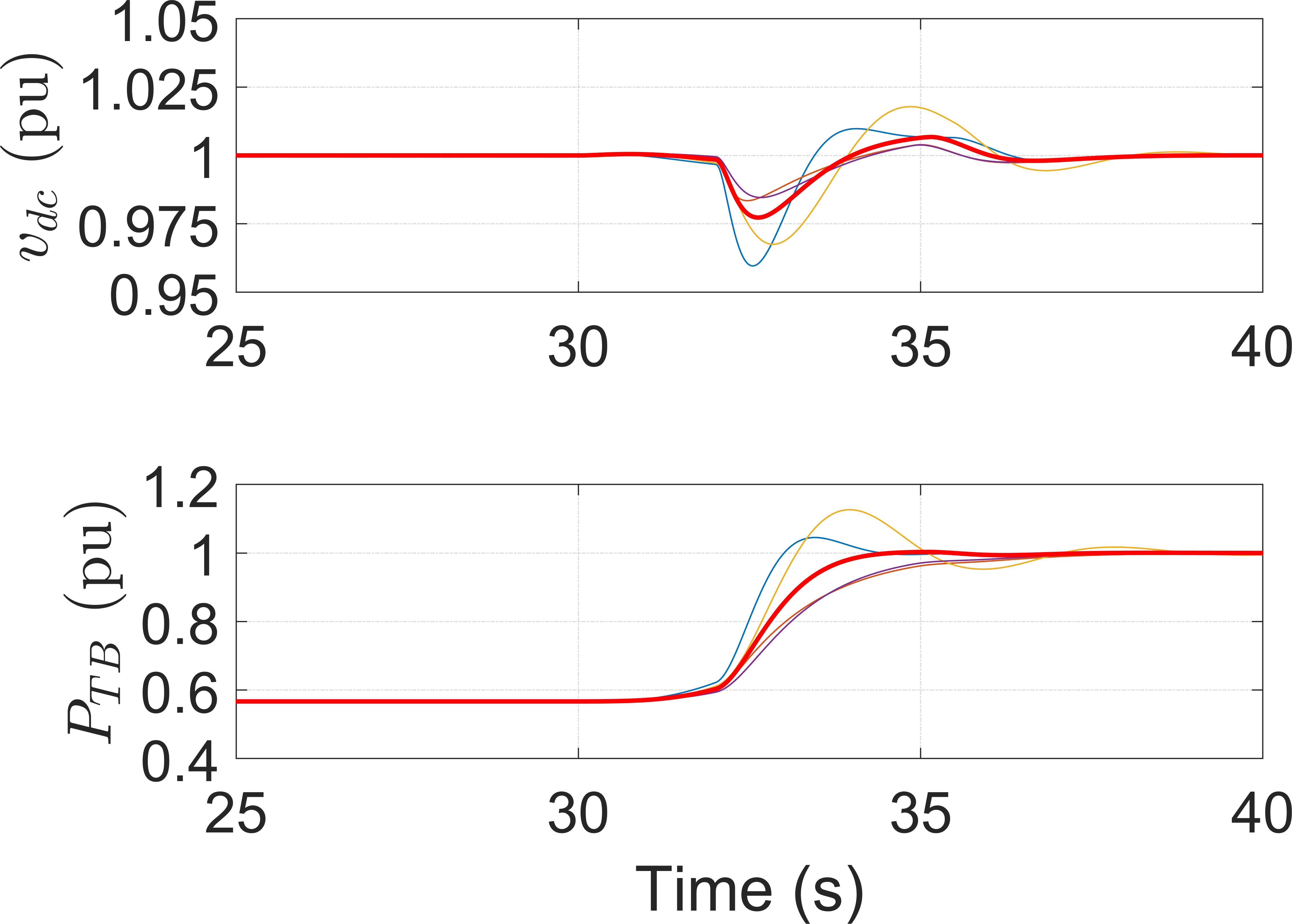}
			}
			\caption{Performance of GFM WTG under optimal and random parameters. The random parameters are chosen in the range given in Table \ref{table:Opt_VSM}. Optimal performance is shown in the red curve.}
			\label{fig:SA_GFM}
			\vspace{-10pt}
		\end{figure}
		
	}

	\subsection{Normal Operation}
	
	\begin{figure*}[!bt]
		\centering
		\subfigure[Region control mode under strong grid (SCR = 10).]
		{
			\includegraphics[width=0.221\linewidth]{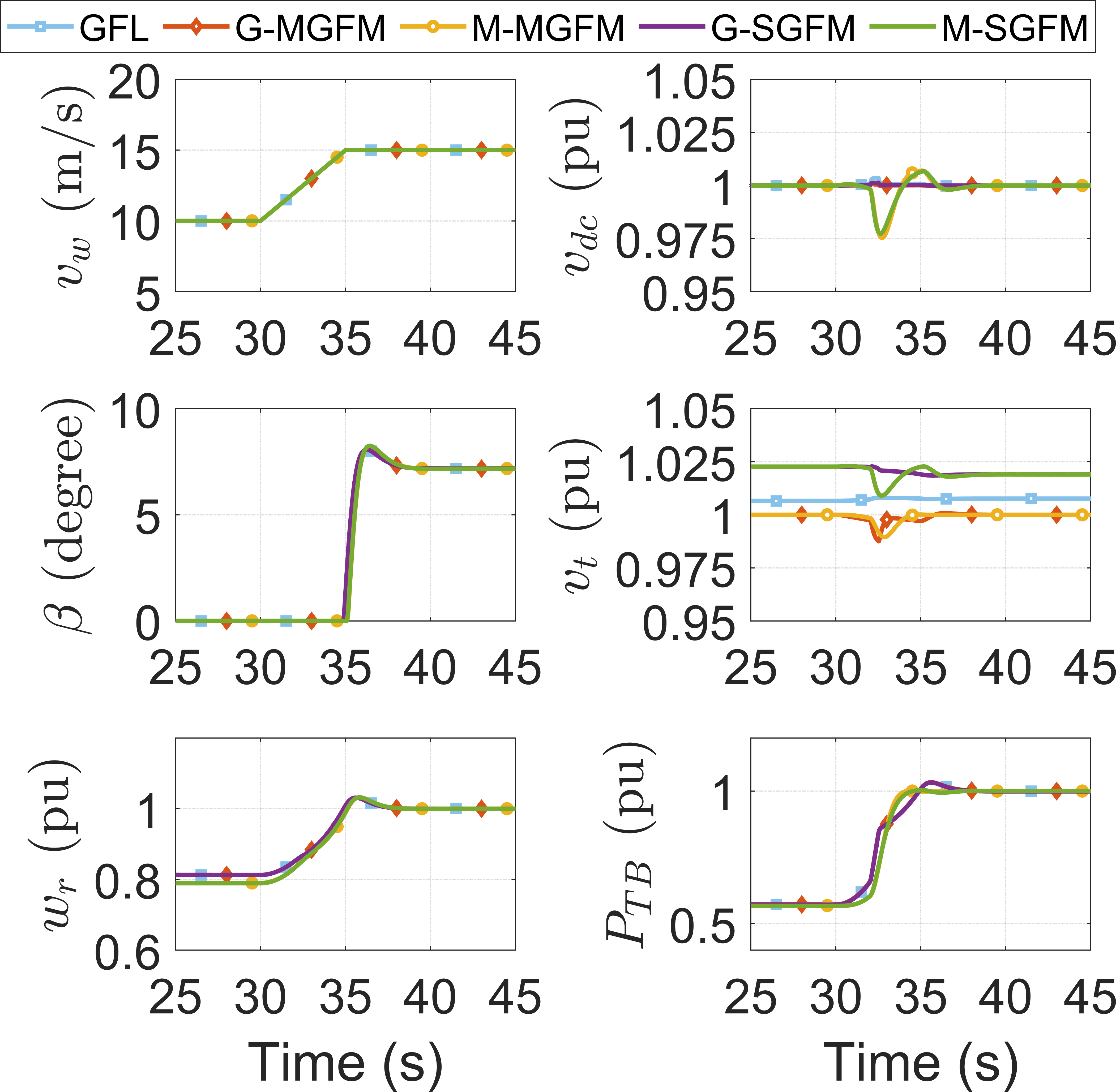}
			\label{fig:dynWind_SCR10}
		} 
		\subfigure[Region control mode under weak grid (SCR = 2.5).]
		{
			\includegraphics[width=0.221\linewidth]{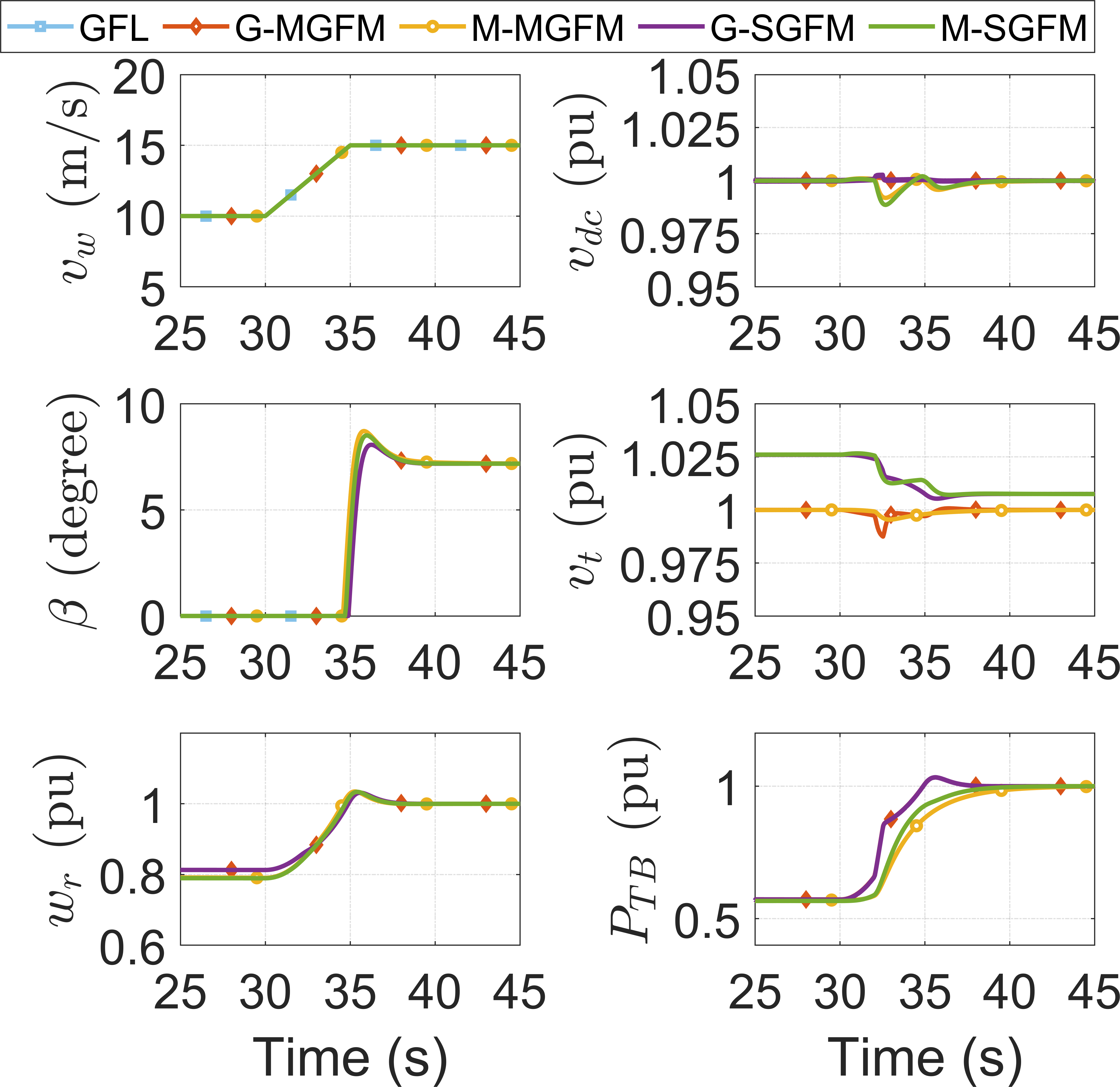}
			\label{fig:dynWind_SCR2p5}
		} 
		\subfigure[Power curtail mode under strong grid (SCR = 10).]
		{
			\includegraphics[width=0.23\linewidth]{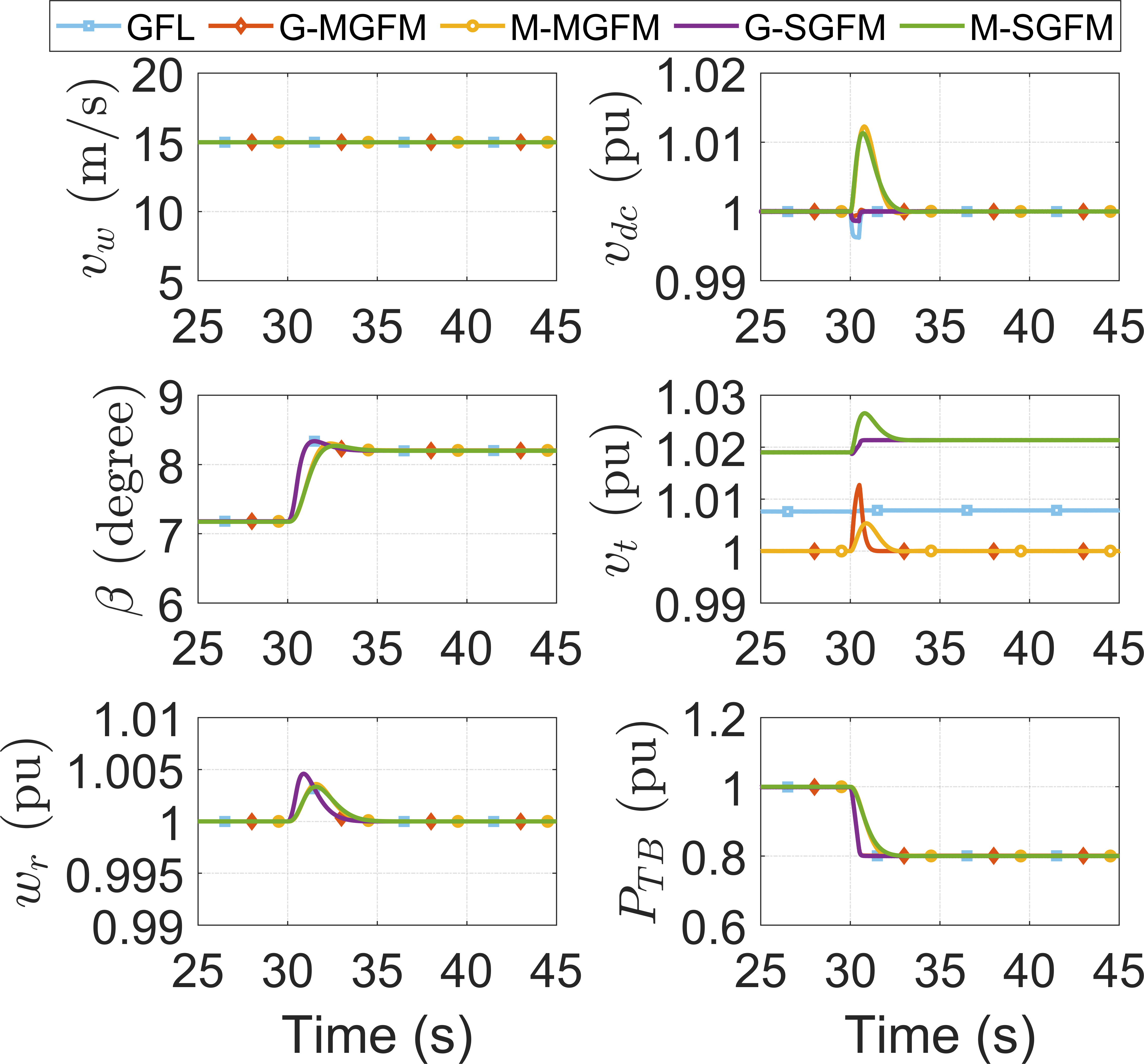}
			\label{fig:Pcurtail_SCR10}
		} %\quad
		\subfigure[Power curtail mode under weak grid (SCR = 2.5).]
		{
			\includegraphics[width=0.23\linewidth]{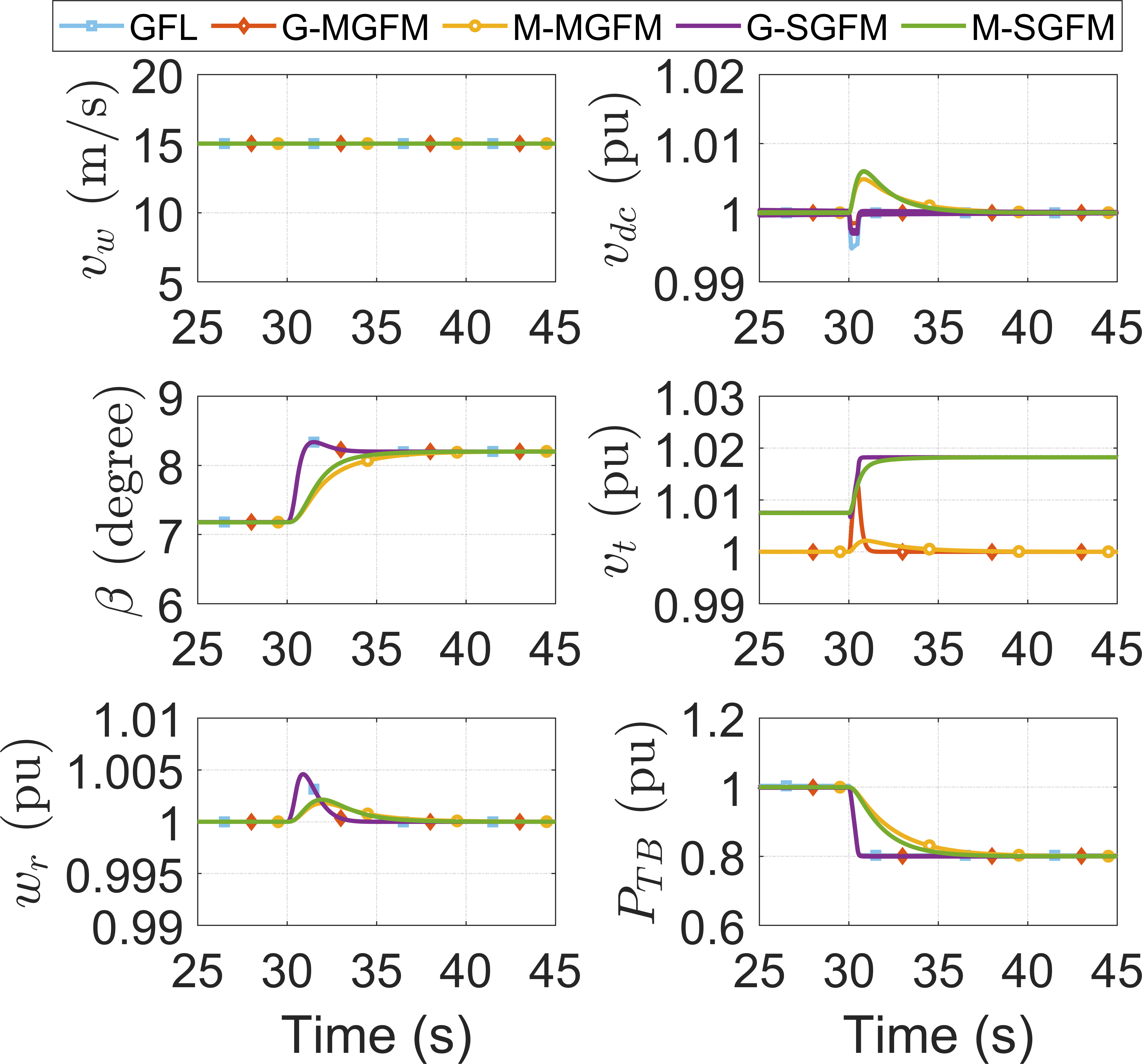}
			\label{fig:Pcurtail_SCR2p5}
		}
		\caption{Performance of GFM WTG under different operation modes. The GFL type fails to maintain stability.}
	\end{figure*}
	
	This section evaluates the performance of the region control function. The wind speed changes from 10~m/s to 15~m/s at 30~s, resulting in WTG operation in regions 2 and 3. The compared result of GFL and GFM controls under strong grid conditions (short-circuit ratio (SCR) = 10) is shown in Fig.~\ref{fig:dynWind_SCR10}. \hlblue{The responses of pitch angle, rotor speed, and output power among these WTGs are similar. With optimized control parameters, the performances of G-GFM types (G-SGFM and G-MGFM) and GFL are identical, except for terminal voltage ($v_t$). There is a slight oscillation in DC-link voltage of M-MGFM and M-SGFM when the controller changes from region 2 to region 3. It can be observed that, with the optimized control parameters, there is no significant discrepancy between the multi-loop (G-MGFM and M-MGFM) and single-loop (G-SGFM and M-SGFM) techniques.}
	
	The comparison of these controllers in weak grid conditions is shown in Fig.~\ref{fig:dynWind_SCR2p5}, in which the SCR value is 2.5. The GFL control fails to keep the WTG in stable operation, whereas all types of GFM controls work normally. \hlblue{Compared to the strong grid condition, the G-GFM types preserve their performance, except for the terminal voltage in the case of G-SGFM type (higher voltage drop than the strong grid case). The M-GFM types show a slower response time in active output power than the strong grid case. Therefore, the control parameters of M-GFM types should be designed with consideration of the grid conditions to achieve better performance.}
	
	The power set-point of WTG is changed from 1~pu to 0.8~pu to evaluate the performance of power curtailment mode. In this mode, the pitch controller regulates the rotor speed of PMSG constantly at 1~pu. The G-GFM types regulate the output power of WTG through the torque controller, whereas the M-GFM types directly regulate output power. Fig.~\ref{fig:Pcurtail_SCR10} shows the comparison of five controllers in strong grid conditions. \hlblue{The advantages of the G-GFM types are the fast response of output power and good tracking performance of DC-link voltage. In the cases of M-GFM types, the output power response is slower than the G-GFM types as it is directly regulated via the VSM scheme of the grid-side controller. The comparison in the condition of the weak grid is shown in Fig.~\ref{fig:Pcurtail_SCR2p5}. The GFL type fails to control WTG in this condition. In comparison to the strong grid state, the G-GFM types exhibit similar performance; but the M-GFM types show a slower response.}
	
	\subsection{Abnormal Operation}	
	%	\begin{figure}[!bt]
	%		\centering
	%		\includegraphics[width=0.84\linewidth]{fault_SCR10.png}
	%		\caption{Fault performance under strong grid. The G-MGFM and M-MGFM types tend to instability.}
	%		\label{fig:fault_SCR10}
	%	\end{figure}
	%	\begin{figure}[!bt]
	%		\centering
	%		\includegraphics[width=0.84\linewidth]{fault_SCR2p5.png}
	%		\caption{Fault performance under weak grid. The GFL, G-MGFM, and M-MGFM types tend to instability.}
	%		\label{fig:fault_SCR2p5}
	%	\end{figure}
	
	\begin{figure}[!bt]
		\centering
		\subfigure[Strong grid (SCR = 10).]
		{
			\includegraphics[width=0.454\linewidth]{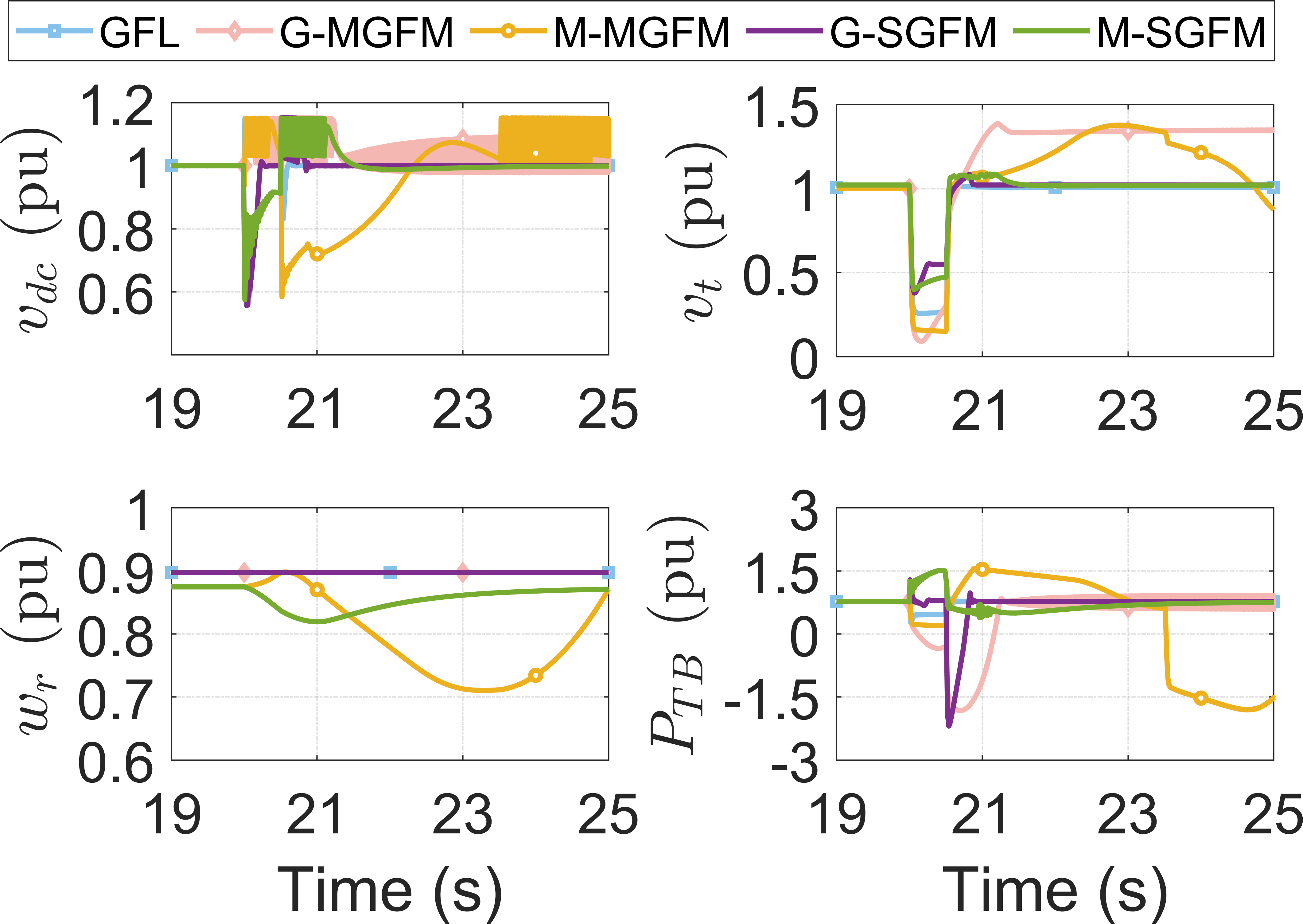}
			\label{fig:fault_SCR10}
		} 
		\subfigure[Weak grid (SCR = 2.5).]
		{
			\includegraphics[width=0.46\linewidth]{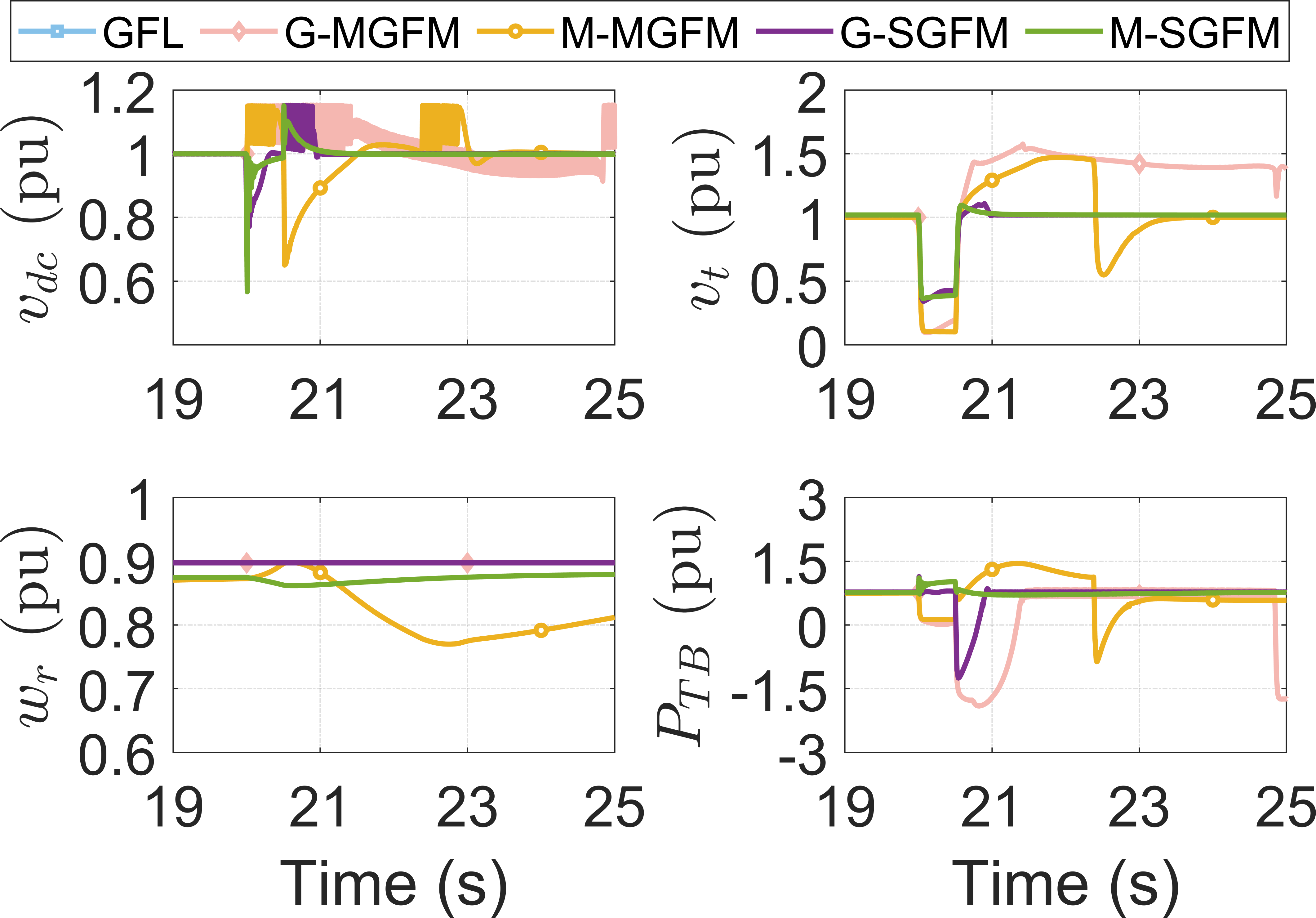}
			\label{fig:fault_SCR2p5}
		} 
		\caption{Performance of GFM WTG under fault condition.}
	\end{figure}
	\begin{figure}[!bt]
		\centering
		\includegraphics[width=0.9\linewidth]{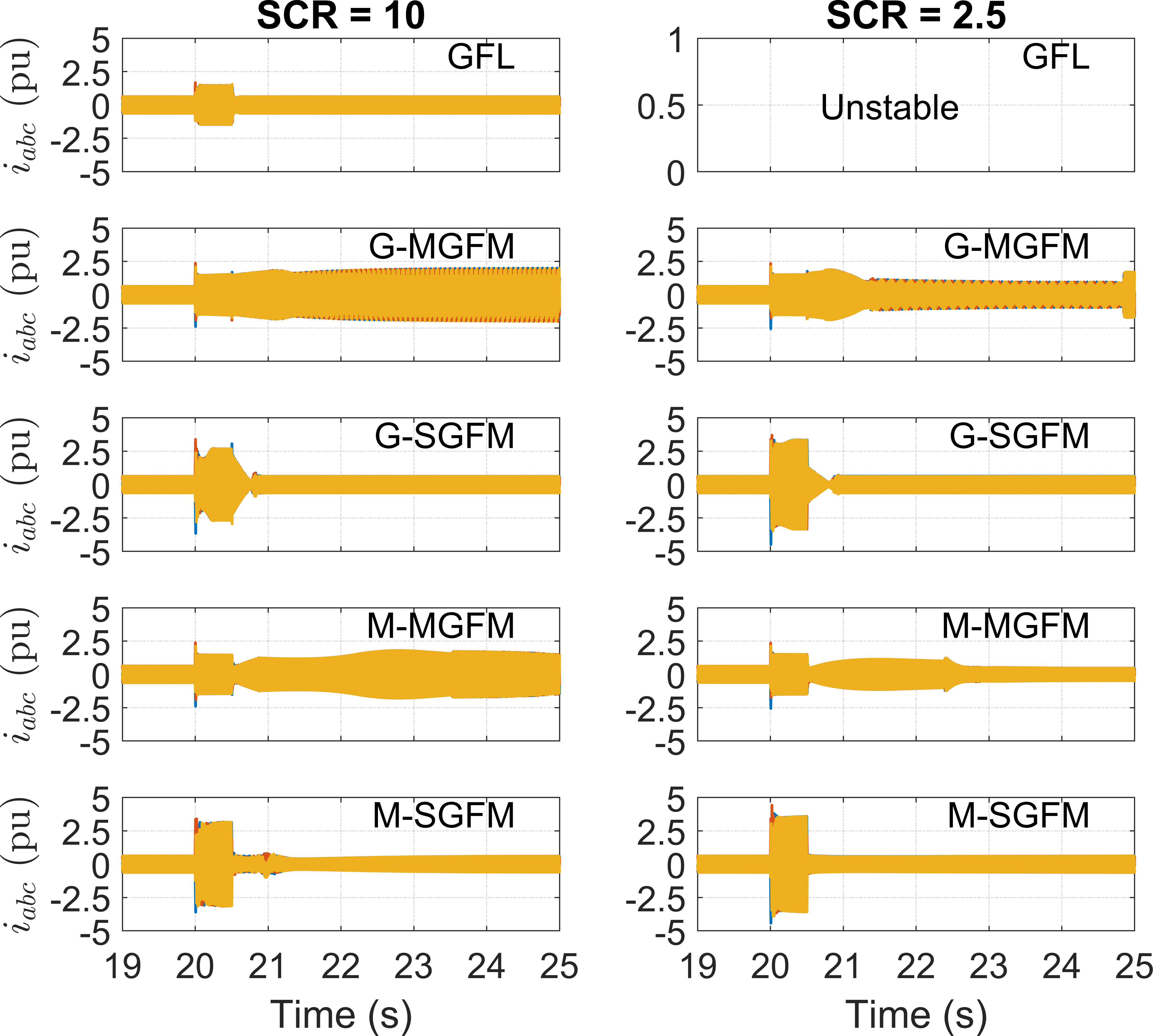}
		\caption{Three-phase transient current of WTG under fault condition.}
		\label{fig:fault_current_RF10}
	\end{figure}
	
	\hlblue{
		One of the biggest challenges of GFM WTGs is the fault current limiting strategy. The current reference saturation strategies are used in the MGFM types to limit the fault current, but it causes instability due to significant phase shift \cite{6872529}. For the SGFM type, the overload mitigation control scheme to limit the fault current is used; however, it is designed to mitigate long-term overload events such as the large load step and loss of generation units. Currently, there are no mature current limiting strategies reported in the literature. This paper examines the transient performance of the GFM WTGs under strategies of current reference saturation and overload mitigation. The GFL, G-MGFM, and M-MGFM types use the current saturation scheme in \cite{7390315} with a maximum current of 1.2~pu, and the G-SGFM and M-SGFM types use the overload mitigation scheme in \cite{8274300} with a maximum power of 1.2~pu.} 
	
	Performance of GFL and GFM controls of WTG under three-phase-to-ground fault is shown in Figs.~\ref{fig:fault_SCR10} and \ref{fig:fault_SCR2p5}. For the typical GFL control, the DC-link voltage rapidly increases when the fault occurs because the GFL control slowly responds to the disturbance. When the DC-link voltage increases significantly, the DC-chopper is activated to protect the DC system. By comparison, because the GFM WTGs respond quickly to the disturbance while the electric power from PMSG changes slowly, DC power from the DC link is drawn to compensate for the disturbance, resulting in the rapid drop of DC-link voltage. This phenomenon causes a negative impact on the control performance of GFM WTG. In addition, the current saturation in multi-loop GFM types causes the phase shift, making MGFM types difficult to recover after clearing the fault. It can be seen that the GFM controls under strong grid conditions tend to lose synchronism easier than the weak grid condition as a small change in phase angle causes significant power oscillation. The M-GFM types gain the advantage in this condition as the machine-side converter controls the DC-link voltage. 
	
	Fig.~\ref{fig:fault_current_RF10} shows the three-phase current of GFL and GFM controls. It can be seen that the GFL control successfully limits the transient current at 1.2~pu when the grid is strong, but it fails to operate in the weak grid condition. \hlblue{The multi-loop GFM controls (M-MGFM and G-MGFM) successfully limit the fault current during the fault, but they failed to recover after clearing the fault. It should be noted that the current saturation scheme of GFL and multi-loop GFM types are the same. It is also observed that the transient current of single-loop controls (M-SGFM and G-SGFM) is much higher than that of the multi-loop controls (M-MGFM and G-MGFM).} After clearing the fault, the multi-loop GFM types lose synchronization due to the current saturation. By comparison, the SGFM types still synchronize with the grid during fault by the overload control scheme, which helps WTG restore to the normal operation.

	\hlblue{
		\subsection{Black Start and Autonomous Operation}
		
		Black start and autonomous operation are the two important requirements for the GFM WTGs. A typical black start strategy for all GFM-WTGs is to use either GSC or MSC to energize the DC-link voltage; then, the remanding converter starts to regulate wind turbine power. For the G-GFM types, as the GSC controls the DC-link voltage, an external source is required to start the GSC converter. The G-GFM types also can start without the external source by coordinating the MSC and GSC controllers in the black start condition. By comparison, the black start strategy for the M-GFM types is straightforward. Firstly, the wind turbine starts to capture wind power, and the MSC starts to regulate the DC-link voltage. The GSC is activated once the DC-link voltage is stable to create the system voltage and frequency. If the wind power is sufficient, the M-GFM types can supply power for the load in autonomous operation mode. The M-GFM types can achieve the self-black start capability without any external source. 
		
		To the best of the authors' knowledge, there are no existing studies on black start and autonomous operation of the G-GFM types. Therefore, only M-GFM types are examined in this paper. Fig. \ref{fig:BS_MGFM} shows the performance of black start and autonomous operation of two M-GFM types. The DC-link capacitor is charged from 0~s. As shown in the top sub-figure, it takes 0.1~s to increase the DC-link voltage from 0 to 1~pu. At 20~s, the grid-side converter is started to form the system voltage and frequency. It can be seen in the bottom figure that the output voltage ($v_t$) rapidly increases from 0 to 1~pu. At 25~s, an $RL$ load is connected to the wind turbine. The output active and reactive power of wind turbines rapidly increases at 25~s. The wind turbine operates under power curtailment mode in the autonomous operation mode. The pitch controller regulates the pitch angle to maintain the rated rotor speed ($\omega_r$) in this operation. It should be noted that the wind power is assumed to be sufficient to supply the load. Further studies on the autonomous operation of GFM WTG in the condition of insufficient wind power are required.
		\begin{figure}[!bt]
			\centering
			\includegraphics[width=0.9\linewidth]{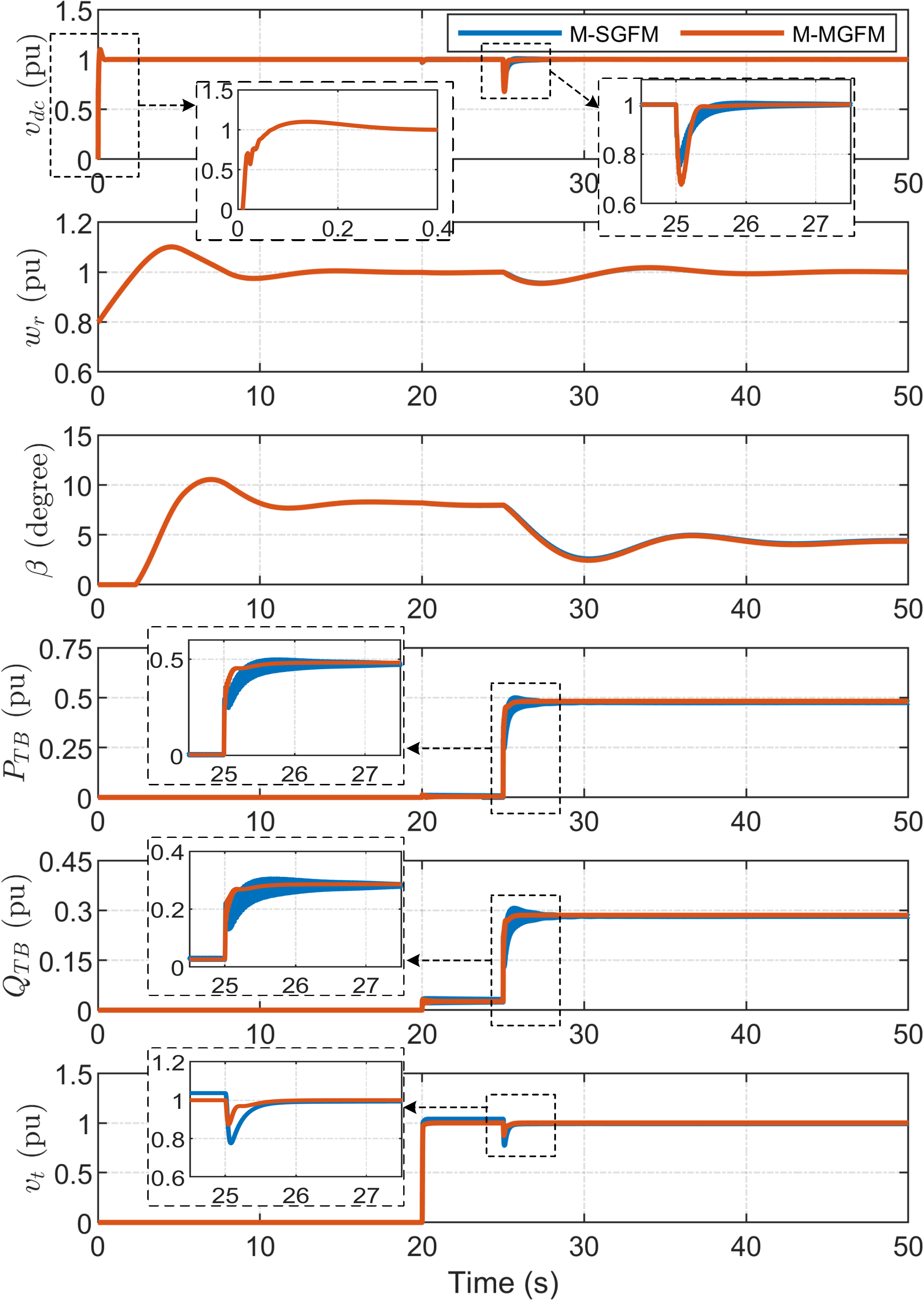}
			\caption{\hlblue{Black start and autonomous operation of M-GFM WTGs.}}
			\label{fig:BS_MGFM}
		\end{figure}
	}

	%	\begin{figure}[!bt]
	%		\centering
	%		\includegraphics[width=0.8\linewidth]{vi_BS_z12.png}
	%		\caption{Voltage and current under RL load condition.}
	%		\label{fig:fault_current_RF10}
	%	\end{figure}

	%	\begin{figure}[!bt]
	%		\centering
	%		\includegraphics[width=0.8\linewidth]{vi_BS_Rload_z2.png}
	%		\caption{Voltage and current under R load condition.}
	%		\label{fig:fault_current_RF10}
	%	\end{figure}

	\vspace{10pt}
	
	\section{Recommendations and Future Directions}
	\label{sec:Future}
	\hlblue{Based on the observations a comparative analysis in the previous sections, this sections provides the recommendations and future directions for 1) advanced controls and protection, and 2) standards development with modeling, energy storage integration, and testings as follows:}
	\hlblue{\subsection{Controls and protection}}
	
	\subsubsection{Stability of GFM WTG}
	
	The detailed comparisons between multi-loop and single-loop GFM inverters in microgrids \cite{8844715} revealed that the stability margin of the single-loop control is larger than the multi-loop control; however, the stability issues of GFM WTGs have not been evaluated. Our paper performed fault studies to evaluate the stability margin of different GFM control methodologies for WTGs. It also revealed that the single-loop GFM WTGs provide a larger stability margin than the multi-loop control WTG do. Figs.~\ref{fig:fault_SCR10} and \ref{fig:fault_SCR2p5} depicted that the MGFM types tend to instability while the SGFM types are stable. The stability margin of the GFM WTG depends on the GFM control structure and the regulation strategies of DC-link voltage. Specifically, M-GFM types can stably restore to the normal operation after clearing the fault, whereas the G-GFM types experience severe disturbance or even instability. The M-GFM types provide a better performance as the machine-side converter controls the DC-link voltage which is decoupled from the grid's disturbance. However, the active power response of M-GFM types under normal operation relies on the grid's strength. Thus, an optimal design of virtual synchronous machines for the M-GFM types should be considered. 
	
	\subsubsection{Current Limiting Strategy}	
	Due to the lack of the current control loop, the SGFM types face the challenge of limiting the transient current. As depicted in Fig.~\ref{fig:fault_current_RF10}, although the SGFM types stably recover in the post-fault condition with an overload mitigation control scheme, the transient current of SGFM types during fault reaches 5~pu, which can damage the power converter. Since existing current saturation scheme of GFL control cannot be used for the MGFM types, different current limiting strategies have been developed, such as switching control strategy from grid-forming to grid-following during fault period \cite{rokrok2021transient}, virtual active power \cite{kkuni2021effects}, using virtual impedance or admittance in addition to current saturation algorithm \cite{qoria2020current, rosso2021implementation}, current reference limiting with anti-windup integration \cite{zhao2020freezing}, current-constrained unified virtual oscillator control \cite{awal2021transient}, and circular current limiter \cite{liu2021optimal}. Most existing current limiting strategies have been developed for the multi-loop GFM controls. As the single-loop GFM controls provide a better performance, further research should be conducted on the current limiter for these GFM controls. In addition, existing current limiting strategies have been developed without considering the dynamic behavior of DC-link voltage during the disturbance, which might be inappropriate for wind turbine applications as the DC-link voltage of WTG varies significantly during fault conditions. A detailed analysis of these current limiting strategies on the full WTG model is essential to verify their feasibility to wind turbine applications. 
	
	\subsubsection{Fault Ride-Through Capability}
	Various open issues of GFM inverters, such as grid synchronization, current limiting, and fault ride-through, have been discussed in \cite{9408354, 9513281}, which persist for the wind turbine applications. Critical clearing time is an essential factor of GFM controls to maintain grid synchronization in post-fault condition \cite{qoria2019critical, rokrok2021transient, 9541070, qoria2020current}. The fault ride-through strategy should be developed in conjunction with current limiting \cite{rosso2021implementation, awal2021transient}. Developing the voltage ride-through strategy in accordance with the international standards is crucial for practical studies of GFM WTGs.

	\subsubsection{Sub-Synchronous and Harmonic Resonance}
	
	Subsynchronous control interaction (SSCI) of Type-3 WTG, which is caused by interaction between controllers of WTGs and series compensated line, has been reported in \cite{8779671, xiesubsynchronous}. \hlblue{The subsubchronous control interaction of Type-4 WTG has been observed in non-compensated weak system \cite{7878693}.} Harmonic resonance is caused by interaction among WTG's controllers and shunt capacitance of offshore transmission cable or shunt capacitors, which has been reported in Toronto system \cite{moharana2012subsynchronous}. The resonance frequency of SSCI is typically between 1 and 10~Hz, whereas the frequency of harmonic resonance is in second or third order frequency. These resonance issues have been observed in the GFL control of WTG. A recent study in \cite{9022884} demonstrated that the single-loop GFM control also experienced the SSCI issue. More detailed analysis of SSCI and harmonic resonance of GFM WTG should be conducted. 
	
	\subsubsection{Wind turbine structural loads}
	
	Electro-mechanical interactions, in which the grid disturbances cause the structural vibration of wind turbines, have been investigated for the GFL wind turbines \cite{edrah2020effects}. The grid-code requirements pose challenges for the design of the mechanical structure and the electrical system of wind turbines \cite{hansen2010grid, hansen2011impact}. The control systems of wind turbines can be appropriately designed to mitigate such interactions \cite{edrah2022electromechanical}. The issue of electro-mechanical interactions is even worse in the case of GFM WTG, especially offshore floating GFM wind turbines. Phase jumps are the most stressful events for the GFM WTG, which results in the rapid increase of current in the GFM WTG and large torque/power oscillations \cite{lundchallenges}. Such oscillations have a severe impact on the wind turbine structural loads. It is required to investigate the impacts of the GFM control methodologies on the wind turbine structural loads.
	
	\subsubsection{Protection of GFM WTG-Based Wind Farms}
	The fault current contribution from the inverter-based resources (IBRs) is limited due to the hardware constraints, which poses a challenge to the protection systems. The lack of negative sequence current injection is another challenge as the protection systems rely on negative sequence components to detect the unbalanced fault \cite{9449692}. The German grid codes require the ability to inject negative sequence component \cite{vde4120}. Although existing multi-loop GFM controls lack the ability to inject the negative sequence component, the sequence-based control strategy can be used to control the negative sequence component \cite{8579187, awal_rachi_yu_husain_lukic_2021}. By comparison, the single-loop GFM types can supply both negative and zero sequence components, which is experimental verified in \cite{gurule2020experimental}.
	
	\hlblue{\subsection{Standards, modeling, integrated storage, and testings}}
	\subsubsection{Technical Standards} 
	Existing IEEE 1547 standard \cite{8332112} was developed for the GFL inverters, which emphasizes reactive power limits, current harmonic limits, and anti-islanding features \cite{lin2020research}. However, these requirements need to be revisited to apply for the GFM inverters which have distinct behaviors from the GFL inverters. For example, as the GFM inverters have the ability to operate in the stand-alone mode due to their voltage-source behavior, the requirement of the anti-islanding function is obsolete. \hlblue{IEEE P2800 standard, which has been approved recently, establishes the capability and performance requirements for the interconnection of large-scale IBRs; but it is not intended to limit the adoption of inverter's technology and controls. The grid-forming inverter was introduced briefly in this standard for informative purpose~\cite{9641506}}. The National Electric Reliability Corporation (NERC) standard PRC-024-2 is used for protective relay settings, which ensures the voltage and frequency ride-through capabilities of generating units \cite{PRC024-2}. However, the frequency limits in this standard, which were based on the limitations of synchronous machines, might not be a suitable metric for tripping IBRs. NERC issued the performance guideline for IBRs \cite{guideline2018bps}, which indicated that the voltage and frequency trip settings should be set as wide as possible. On the other hand, a Grid-code technical specification for GFM inverters has been proposed by National Grid ESO in \cite{NationalGridESOGFM}, which emphasizes inertia active power, fast fault current injection, and phase jump. These issues have also been discussed by the European Network of Transmission System Operators for Electricity \cite{christensen2020high}. However, these requirements should account for the operating condition of wind generators because the WTGs operating at low power have a limited ability to respond. The fast fault current injection requirement should consider its impact on DC-link voltage regulation to ensure that the WTG can restore to normal operation after disturbance. The challenges and opportunities of GFM IBRs have been discussed by Western Electricity Coordinating Council (WECC), and Energy Systems Integration Group (ESIG) \cite{ESIGGFM}.
	
	\subsubsection{Modeling of GFM WTG for Large-Scale System studies} The generic wind turbine models have been developed by Western Electricity Coordinating Council (WECC) \cite{remtf2014wecc} and International Electrotechnical Commission (IEC) \cite{8488530} for transient studies of large-scale wind farms. However, these models were developed for the GFL WTG. The generic WECC model has been improved to represent the behavior of GFM inverters \cite{9447546}. \hlblue{Recently, WECC approved a generic droop-based GFM IBR model \cite{du2021model}.} Phasor-approximation model has been developed in \cite{misyris2021grid} for RMS modeling of GFM inverters. Three-phase electromechanical models of GFL and GFM inverters have been developed in \cite{du2020modeling} for dynamic simulation of the large-scale unbalanced distribution system. Nevertheless, the standards-compliant model is required for electromagnetic transient studies of large-scale wind farms. 
	
	\subsubsection{Integrated Energy Storage System}
	ESS integrated into WTG would release the grid-side converter from the main wind turbine controllers. Thus, most existing GFM methodologies can be applied for the grid-side converter as the DC-link voltage is controlled by the integrated ESS \cite{6339035}. In addition, the integrated ESS would play the role of energy buffer, which is an important factor to address the uncertain characteristic of wind power \cite{6359915, 9420057}. Practical experience of GFM WTG in \cite{roscoe2019practical} has revealed that the turbine's ability to respond to the disturbance may be affected if the wind speed is declining or the WTG operates at low power. The integrated ESS would help the GFM WTG have the ability to respond to the grid disturbance in such conditions. 
	
	\subsubsection{Testing and Validating GFM WTG}
	The successful testing of a 69MW wind park operating in GFM mode demonstrates GFM wind turbines' feasibility ~\cite{roscoe2020response, roscoe2019practical}. However, there is a lack of explicit criteria for the functionality and performance of GFM WTG. Standardized guidelines on the testing and validating are needed to assess the GFM controls of WTG, which should be clearly defined in standards or grid codes. 
	
	\vspace{10pt}
	\section{Conclusion}
	\label{sec:Conclusion}
	
	This paper provided an overview of grid-forming wind turbine generators and discusses the implementations of different GFM categories.  Four GFM methodologies were discussed: multi-loop GFM (S-MGFM and M-MGFM) and single-loop GFM (S-SGFM and M-SGFM). A comparative study has been conducted to evaluate their performances. It has been observed that the M-MGFM and M-SGFM types provide a better performance in fault conditions as the DC-link voltage is controlled by the machine-side converter which is decoupled from the grid's disturbance. The tested result under fault conditions showed that the single-loop GFM types have a higher stability margin than the multi-loop GFM types. This paper revealed that the DC-link voltage has a significant impact on the performance of the GFM WTGs that should be taken into consideration for control design. Finally, recommendations and future trends of GFM WTGs have been provided.

	%	% use section* for acknowledgment
	%	\section*{Acknowledgment}
	%	
	%	
	%	This material is based upon research supported by, or in part by, the New York State Energy Research and Development Authority (NYSERDA) under award number 148516.

	\ifCLASSOPTIONcaptionsoff
	\newpage
	\fi

	\balance % uncomment this can remove two blanks last pages.
	\bibliographystyle{IEEEtran}
	\bibliography{References}
	
	%\EOD
\end{document}